\RequirePackage{fix-cm}
\documentclass[fleqn,usenatbib]{mnras}

\usepackage{newtxtext,newtxmath}

\usepackage[T1]{fontenc}

\DeclareRobustCommand{\VAN}[3]{#2}
\let\VANthebibliography\thebibliography
\def\thebibliography{\DeclareRobustCommand{\VAN}[3]{##3}\VANthebibliography}

\usepackage{graphicx}	
\usepackage{amsmath,tensor,accents}	
\usepackage{xspace}
\usepackage{booktabs}
\usepackage[table]{xcolor} 
\usepackage[export]{adjustbox}
\usepackage{comment}
\usepackage{bm}
\usepackage{float}
\usepackage[normalem]{ulem}

\graphicspath{{figures/in_paper/}}

\newcommand{\psib}{$\Psi_\mathrm{B}$\xspace}
\newcommand{\barpsib}{$\bar{\Psi}_\mathrm{B}$\xspace}
\newcommand{\tildepsib}{$\widetilde{\Psi}_\mathrm{B}$\xspace}

\newcommand{\rg}{$r_\mathrm{g}$\xspace}
\newcommand{\rgc}{$r_\mathrm{g}/c$\xspace}

\newcommand{\bhac}{{\tt BHAC}\xspace}

\newcommand\carteq{\mathrel{\overset{\makebox[0pt]{\mbox{\normalfont\tiny\sffamily Cart.}}}{=\joinrel=}}}
\newcommand\mkseq{\mathrel{\overset{\makebox[0pt]{\mbox{\normalfont\tiny\sffamily MKS}}}{=\joinrel=}}}

\newcommand\thickbar[1]{\accentset{\rule{.4em}{.8pt}}{#1}}

\title[Plasmoid identification in MHD simulations]{Plasmoid identification and statistics in two-dimensional Harris sheet and GRMHD simulations}

\author[J.T. Vos et al.]{
  J.T. Vos,$^{1}$\thanks{E-mail: jt.vos@astro.ru.nl}
	H. Olivares,$^{1}$
    B. Cerutti$^{2}$ and
    M.A. Mo\'scibrodzka$^{1}$
\\
$^{1}$Department of Astrophysics/IMAPP, Radboud University, PO Box 9010, 6500 GL Nijmegen, the Netherlands\\
$^{2}$Univ. Grenoble Alpes, CNRS, IPAG, 38000 Grenoble, France \\
}

\date{Accepted XXX. Received YYY; in original form ZZZ}

\pubyear{2023}

\begin{document}
\label{firstpage}
\pagerange{\pageref{firstpage}--\pageref{lastpage}}
\maketitle

\defcitealias{uzdensky10}{ULS}

\begin{abstract}
Magnetic reconnection is a ubiquitous phenomenon for magnetized plasma and leads to the rapid reconfiguration of magnetic field lines.
During reconnection events, plasma is heated and accelerated until the magnetic field lines enclose and capture the plasma within a circular configuration.
These plasmoids could therefore observationally manifest themselves as hot spots that are associated with flaring behaviour in supermassive black hole systems, such as Sagittarius A*.
We have developed a novel algorithm for identifying plasmoid structures, which incorporates watershed and custom closed contouring steps.
From the identified plasmoids, we determine the plasma characteristics and energetics in magnetohydrodynamical simulations.
The algorithm’s performance is showcased for a high-resolution suite of axisymmetric ideal and resistive magnetohydrodynamical simulations of turbulent accretion discs surrounding a supermassive black hole.
For validation purposes, we also evaluate several Harris current sheets that are well-investigated in the literature.
Interestingly, we recover the characteristic power-law distribution of plasmoid sizes for both the black hole and Harris sheet simulations.
This indicates that while the dynamics are vastly different, with different
dominant plasma instabilities, the plasmoid creation behaviour is similar. Plasmoid occurrence rates for resistive general relativistic magnetohydrodynamical simulations are significantly higher than for their ideal counterpart.
Moreover, the largest identified plasmoids are consistent with sizes typically assumed for semi-analytical interpretation of observations.
We recover a positive correlation between the plasmoid formation rate and a decrease in black-hole-horizon-penetrating magnetic flux.
These results demonstrate the efficacy of the newly developed algorithm which has enabled an extensive quantitative analysis of plasmoid formation for black hole accretion simulations.
\end{abstract}




\section{Introduction}
Flaring events, at the X-ray and infrared wavelengths, are known to occur on a daily basis for the supermassive black hole (SMBH) at the center of the Milky Way, Sagittarius A$^\ast$ 
(hereafter Sgr A$^\ast$, \citealt{baganoff01,genzel03,eckart04,witzel21}).
The SMBH has an estimated mass of $M \approx 4 \times 10^6 \: M_\odot$ and lies at a distance of $D \approx 8$ kpc as was established by long-term monitoring programs of the source and dynamics of orbiting stars \citep{ghez08,gillessen09a,gillessen09b,gillessen17,gravity_s2_18,gravity19,do19b}.
At sub-mm / mm wavelengths, Sgr A* is known to be a stochastically ($\mathcal{O}(10\%)$ over hours) variable source that is associated with the (stereotypical) \emph{quiescent} accretion state. 
While flares at NIR/X-ray wavelengths correspond to significant increases in flux, ``\emph{flaring}'' events at mm-wavelenghts are typically hard to disentangle from the background variability \citep{eht22sgrai,wielgus22}. 
Recently, it was shown that mm-wavelength light curves observed with the Atacama Large Millimeter/submillimeter Array suggest orbital motion of a hotspot quickly after an X-ray flare \citep{wielgus22b}.
Previously, this was also established in the NIR band \citep{gravity_plasmoids_18}.
The physical mechanism that causes these flares is currently not well-understood, but a number of working theories associate them with strongly magnetized anisotropies in the accretion flow \citep{broderick05,broderick06,gravity_jimenez20,dexter20,porth21,vos22,vos23timelags,ripperda22}.

One such scenario that may explain these flares and the formation of hot spots is the formation of plasmoids as part of a magnetic reconnection event \citep[e.g.,][]{ripperda20,ripperda22,elmellah23}.
This is a phenomenon that occurs in a vast number of astrophysical sources, including pulsar wind nebulae, magnetars, black hole and neutron star magnetospheres, or relativistic jets of active galactic nuclei \citep{kagan15}.
Magnetic reconnection \citep[][for a review]{uzdensky22} can broadly be thought of as a rapid reconfiguration of the magnetic field geometry at the interface of opposite polarity magnetic fields that results
in the formation of a magnetic island with a typical circular magnetic field morphology.
After the closing of the magnetic field lines, plasma is trapped within the magnetic field structure, creating what is known as a plasmoid.
The reconfiguration is often accompanied by particle acceleration to high (non-thermal) energies \citep{werner18} - effectively converting electromagnetic energy into particle kinetic energy (thermal and non-thermal).
A theoretical description for the large-scale dynamics of magnetic reconnection in idealized configurations was established by \citet{sweet58} and \citet{parker57}.
This picture is, however, too simplistic for our purposes as it does not deal with plasmoid formation.
To model the plasmoid-unstable regime, one has to adopt a numerical approach via particle-in-cell (PIC) or magnetohydrodynamical (MHD) simulations. Both methods will be outlined in detail in the following paragraphs.


Fully kinetic PIC methods generally assume a collisionless description that consists of ion-electron, electron-positron (pair), or ion-pair plasma \citep[][for a review]{kagan15}.
These methods are considered first principle as they naturally impose both a spatial (``skin depth'' $c/\omega_p$) and temporal ($\omega_p^{-1}$) scale via the plasma oscillation frequency $\omega_p= {\scriptstyle \sqrt{4\pi n q^2 / w_n m}}$, where $n$, $q$, $m$, $w_n$ are the particle number density, charge, mass, and enthalpy, respectively.
While MHD methods only describe the plasma's bulk motion and characteristics, PIC methods track the velocities, trajectories, and energies of individual particles.
Collisionless plasma studies have been conducted to investigate various physical scenarios; 
isolated (Harris) current sheets in 2D \citep{cerutti12,nalewajko15,kagan16,sironi16,petropoulou18} and 3D \citep{sironi14,cerutti14a,guo16,werner17},
configurations investigating magnetic turbulence \citep{comisso19,borgogno22,bacchini22},
and (general-relativistic) accretion simulations describing plasma within the magnetosphere of compact objects (for black holes; \citealt{parfrey19,crinquand20,crinquand21,elmellah22,elmellah23}, or neutron stars; \citealt{chen14,philippov18,guepin20,cerutti21}).
Although PIC methods are instrumental in, e.g., understanding the origin of non-thermal emission, they remain confined to microscopic plasma scales, which makes interpretation at astrophysically large scales difficult. 

General relativistic magnetohydrodynamical (GRMHD) methods have been extensively and successfully used to describe the macroscopic picture of accretion onto SMBHs (for M87*; \citealt{eht1,eht5,eht8m87pol21}, for Sgr A*; \citealt{eht22sgrai,eht22sgrav}).
Almost exclusively, one assumes an ideal GRMHD description that is an inadequate framework for capturing magnetic reconnection and the formation of plasmoids \citep{ripperda20}, as the plasmoid-instability is triggered due to numerical limits rather than consistently resolving the underlying current sheet.
Resistive GRMHD does give a scale to the current sheet and makes it resolvable \citep[][and reference therein]{ripperda19} by means of imposing a constant resistivity ($\eta$) in the simulations.
While the physical resistivity is likely spatially and temporally variable, a uniform scalar resistivity already helps to consistently capture the dynamics associated with magnetic reconnection in the accretion flow.
Even though not physically or numerically well-constrained, we point out that magnetic reconnection and plasmoid formation does occur in ideal GRMHD, where numerical limits effectively impose the minimally achievable resistivity.

In this work, we investigate plasmoid formation from fast relativistic reconnection for plasmoid-forming astrophysical plasma in both ideal and resistive GRMHD.
To be able to assess the plasmoid formation dynamics, we need to address another, equally important aspect which is that plasmoid structures are difficult to isolate from their surroundings.
Therefore, we have developed a novel analysis algorithm for detecting them. 
It deviates significantly from plasmoid-finding methods employed previously for GRMHD simulations \citep{nathanail20} and is more akin to the methods employed in PIC studies by \citet{sironi16,hakobyan19,hakobyan21}.
However, as in a fluid MHD description one does not have the luxury of individual particle trajectories, we apply our analysis fully in post-processing which gives it more flexibility.
Using our methodology, we investigate the differences in occurrence rate, morphology, size, and typical plasma parameters of plasmoids in both ideal and resistive GRMHD for a newly created suite of 2.5D simulations with exquisite resolution.
To showcase the validity and high fidelity of the algorithm we also apply it to a set of Harris current sheet simulations that are equally well-resolved.

The paper is structured as follows.
An in-depth description of the methods we use to simulate and identify these features are outlined in Section~\ref{sec:methods}.
The results and their interpretation are presented in Section~\ref{sec:results}. The discussion and conclusion can be found in Sections~\ref{disc:discussion} and \ref{conc:conclusion}.

\section{Methods} \label{sec:methods}
In the following sections, we will describe the algorithm that identifies the plasmoids and outline the two setups we investigate.

\subsection{Relativistic MHD primer: ideal and resistive}
The plasma flows of both the Harris sheet and BH accretion disc are simulated within the framework of the Black Hole Accretion Code \citep[{\tt BHAC},][]{porth17,olivares19}, which solves the (resistive; \citealt{ripperda19}) MHD equations in stationary spacetimes. These equations are defined as;
\begin{align}
	&\nabla_\mu (\rho u^\mu) = 0, \\
	&\nabla_\mu T^{\mu\nu} = 0, \\
	&\nabla_\mu \tensor[^\star]{F}{^{\mu\nu}} = 0,
\end{align}
where $\nabla_\mu$ denotes the covariant derivative, $\rho$ the rest-mass density, $u^\mu$ the fluid four-velocity, $T^{\mu\nu}$ the energy-momentum tensor (containing both ideal fluid and electromagnetic fields), and $\tensor[^\star]{F}{^{\mu\nu}}$ the (Hodge) dual of the Faraday tensor.
\bhac is a versatile code that sets the speed of light $c$ to unity and utilises Lorentz-Heaviside units, which effectively incorporates the $\sqrt{4\pi}$ factors into the electro-magnetic quantities.

In this work, we utilize both ideal and resistive  MHD.
The main difference between both these approaches is the way they handle the evolution of the electric field, which is denoted by
\begin{equation}
  \bf{E} = -\bf{v} \times \bf{B} + \eta J.
\end{equation}
Note that the resistivity is denoted by $\eta = 1 / \sigma_c$ where $\sigma_c$ is the conductivity.
While in resistive MHD the electric field ($\bf{E}$) includes an explicit calculation of the resistive Ohm's law to get an expression for $\bf{J}$, in ideal MHD it is inferred directly from the magnetic field (via $\bf{E} = -\bf{v} \times \bf{B}$, also known as the ``frozen-in condition'').
Effectively, one assumes the plasma to be perfectly conducting ($\sigma_c \rightarrow \infty \Rightarrow \eta \rightarrow 0$) in the ideal MHD limit, which is a useful and macroscopically valid approximation in large parts of the accretion disc domain but not when it comes to the formation of plasmoids and other non-ideal effects.
More specifically, the resistivity $\eta$ is not exactly zero in the ideal case (except for infinite resolution), but rather determined numerically by the underlying resolution (or cell size $\Delta x$) which implies that $\eta_\textrm{ide} \propto \Delta x^k$ with $k\approx2$ depending on the accuracy of the fluid evolution scheme \citep{ripperda22}.
The physical interpretation of the resistivity $\eta$ is that it acts as a proxy for kinetic effects within the plasma.

We investigate plasmoid formation from fast relativistic plasmoid-dominated reconnection.
Whether the plasma becomes plasmoid unstable is determined by the Lundquist number $S = L^\prime v_a / \eta$, with typical length of the current sheet $L^\prime$ and the Alfv\'en velocity $v_a$ (see section \ref{meth:harris} for definition).
In order to trigger the fast reconnection and tearing- or plasmoid-unstable regime, the Lundquist number needs to satisfy $S > S_\mathrm{crit}$ where $S_\mathrm{crit} \sim 10^4$ \citep{loureiro07,bhattacharjee09,uzdensky10}.
Note that the Lundquist number is largely determined by the underlying resistivity ($\eta = 5 \cdot 10^{-5}$) which is set as a constant and uniform quantity in our resistive simulations.
Then, if we estimate probable values of $L^\prime \approx 1$ and $v_a \approx c = 1$, we find $S = 2 \times 10^4$ which lies above the threshold.
At first glance, for the ideal simulations, one might think that as $\eta_\textrm{ide}$ is very small it reaches a sufficiently high Lundquist number.
Even though this is the case, the resulting current sheet will always be under-resolved (as it is determined by the underlying resolution) and typically has a width comparable to a singular grid cell \citep{ripperda20}.
This indicates that the tearing-instability is not triggered in the same way as for the resistive simulation and will likely result in differences in plasmoid formation statistics.

\begin{figure*}
    \centering
    \includegraphics[width=\textwidth]{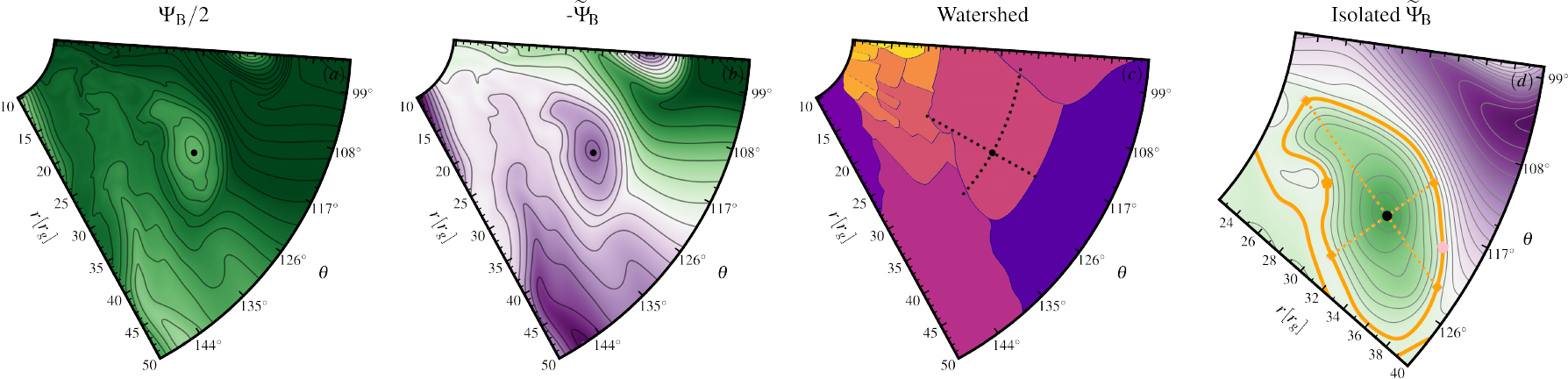}
    \includegraphics[width=\textwidth]{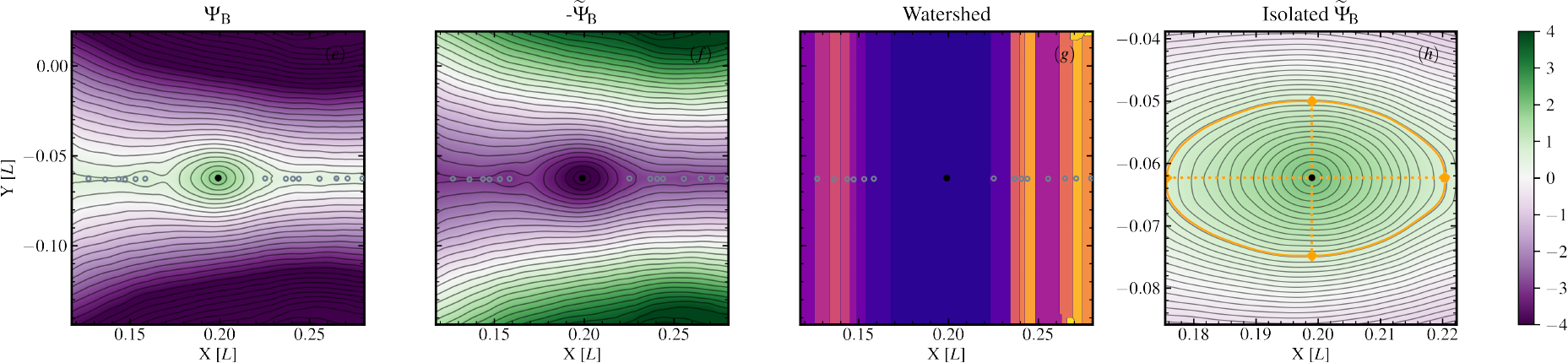}
    \caption{
      A schematic decomposition of the plasmoid identification algorithm. In the \textit{top} panels (a-d), we display a snapshot of a GRMHD simulation ({\tt rM3} at $T=3000$ \rgc) at various points in the pipeline. In the \textit{bottom} panels (e-h), we find the same but for one of the Harris sheet cases ({\tt Hb} at $T=2.93 \, t_c$). 
    In the left column (panels a\&e), one finds the base magnetic flux function \psib -- the starting point.
    To apply the watershed (panels c\&g), one needs to make sure that the plasmoid corresponds to a local minimum which is done with the quantity -\tildepsib (panels b\&g).
    The last column (panels d\&h) showcases how the maximal contour is found for the watershed segment and how the plasmoid's width and height are determined (between the orange diamonds).
    The evaluated O-point is denoted by the black circles. Other O-points in the displayed simulation domain are denoted by the open grey circles.
    }
    \label{fig:watershed_schematic}
\end{figure*}

\subsection{Plasmoid identification} \label{meth:identification}

The starting point of our plasmoid identification routine lies in finding a quantity that lays bare the intrinsically circular magnetic field geometry.
A natural choice for this identification quantity would then fall to the magnetic flux function;
\begin{align}
    \Psi_\mathrm{B} &\mkseq \int \sqrt{-g} B^r \,d\theta, \label{eq:flux_function_mks}  \\
    &\carteq \int B^x \, dy - \int B^y \, dx \label{eq:flux_function_cart}
\end{align}
where $\sqrt{-g}$ is the metric determinant.
Note that $\sqrt{-g} B^r$ corresponds to the magnetic field in the Eulerian frame and that the magnetic flux function \psib corresponds to $A_\phi$, except for a minus sign discrepancy \citep{sironi16}.
The magnetic flux function $\Psi_\mathrm{B}$ is a good choice as its isocontours will follow the inplane magnetic field lines (i.e., $\mathbf{B} \cdot \bm{\nabla} \Psi_\mathrm{B} = 0$).
More specifically, as plasmoids are characterized by their circular magnetic field configuration, the plasmoid center will correspond to the local maxima or minima in $\Psi_\mathrm{B}$ ("O-points").

Due to our methodology, the base \psib structure is not the ideal starting-point of the pipeline.
We, therefore, work with the following quantity;
\begin{equation}
  \widetilde{\Psi}_\mathrm{B} = \bar{\Psi}_\mathrm{B} - \Psi_\mathrm{B}, \label{eq:avg_flux_function}
\end{equation}
where \barpsib scalar denotes the (image) averaged flux function at a given time. 
The removal of the averaged flux function allows for clearer identification of plasmoids in \tildepsib.  

As we now have a suitable medium from which we can start to identifying the plasmoids, we will need a method that is able to classify the magnetic island structure reliably.
For this purpose, we have developed an algorithm that consists of four steps:
\begin{enumerate}
  \item All simulations contain a lot of fine-structure in the magnetic flux function. This makes it hard to differentiate between (magnetic) turbulence and more global features that correspond to a presence of a plasmoid. Therefore, to make certain we filter out much of the turbulence, we apply a blurring (Gaussian or flat) kernel to the flux function (\tildepsib).
  This also gives us control over the size of features we want to be sensitive to.
  The blurring step, however, requires (manual) fine-tuning depending on resolution and nature of the setup.
  Interestingly, to extract the global structure from highly turbulent primary needs the most blurring relatively, while the GRMHD simulation are well-served with a fairly light blurring method.
  \item Following the blurring step, we identify the local minima or maxima that will correspond to the plasmoid's center.
  \item Then, we apply a watershed algorithm \citep[well-described in, e.g.,][]{beucher18} to isolate the domain of interest around the local minimum. We have chosen an implementation that is based on \citet{vincent91}. The watershed segmentation is then used to make an informed cut-out of the domain that will contain a single (local) maxima, so that we have control over what is being fitted while simultaneously improving the quality of the fit.
  \item Lastly, we draw the maximally possible contour within the isolated segment. Utilizing the inherent symmetry in the systems, we sample the space efficiently by means of a binary search from opposite sides (i.e., left and right from center along $\hat{x}$ for the Harris sheet and inner and outer radii along $\hat{r}$ for the GRMHD setups). The resulting contour enables us to gauge the plasmoid's size and orientation, and enables calculations of the plasma quantities associated with the plasmoid and its direct vicinity.
\end{enumerate}

In Fig. \ref{fig:watershed_schematic}, one finds a schematic summary of the points discussed above.
Additionally, it becomes clear that both setups differ fundamentally from one another and therefore warrant a different configuration of the algorithm.
The main differences will be summarized below.
(i) As the Harris sheet setups have periodic boundaries, one needs to be careful to catch plasmoids that are on the boundary.
(ii) Additionally, capturing both "big" and "small" plasmoids in the Harris sheet setup required two different approaches, mainly concerning the blurring kernel. For the big features, one has to apply to a relatively small kernel many times (several hundred times works well in our experience) to not flatten out the global structure too much. To capture the smallest features, one only has to apply the small blurring kernel a few times. One acquires the master set by combining the output from both described configurations following the \citetalias{uzdensky10} criterion.
(iii) For GRMHD, one has to take into account that resolution is concentrated near the black hole and in the equatorial plane and therefore has non-uniform cell-sizes.
(iv) Due to this non-uniform grid layout (for the GRMHD simulations), applying a kernel blur manifests itself differently in various regions of the simulation domain. When applying a relatively small blurring kernel, this effect is minor and manageable. If this is not sufficient, then we interpolate the data to an uniform grid structure.

Lastly, we would like to note more explicitly how plasmoids are identified using the magnetic flux function in other works.
In essence, one identifies plasmoids via the so-called "O"- and "X"-points.
O-points corresponds to the local minima and maxima of the magnetic flux function and denote the center of a plasmoid.
X-points are saddle points and lie in between O-points.
Along a current sheet one therefore expects these points to succeed one another.    
One typically finds the extrema by calculating the Hessian matrix of the magnetic flux function \citep{servidio09,servidio10,zhdankin13,kadowaki18,zhou20} via;
\begin{equation}
H^{\Psi_\mathrm{B}}_{ij}({\bf{x}}) = \frac{\partial^2 \Psi_\mathrm{B} ({\bf{x}})}{\partial x_i \partial x_j} \, .
\end{equation}
Then, one calculates the matrix determinant of the Hessian ($|H^{\Psi_\mathrm{B}}|$) to find the critical points that correspond to $|H^{\Psi_\mathrm{B}}({\bf x})| = 0 \,$ at a given coordinate ${\bf x}$.
The eigenvalues of the Hessian then determined if we have an O-point if it is a local minima (positive definite Hessian) or maxima (negative definite Hessian).
For an X-point, one finds both positive and negative eigenvalues of the Hessian \citep{servidio10}.
However, in our methodology, there is no need to explicitly calculate the Hessian to identify the O- and X-points as these are naturally picked up by the watershed algorithm.
The X-points, which are typically harder to identify \citep{zhdankin13}, will lie on the border of a watershed segment. 
For the O-points, we straight-forwardly calculate the local extrema in a segment.
Finding the critical points in these turbulent maps is a complicated endeavour, as is also illustrated by the computationally intensive mitigation techniques employed in \citet{servidio10}.
Our methodology works around this problem in a relatively natural manner, but this implies that we do not know the exact orientation of the current sheet as the X-point locations are not calculated (other than being on the watershed segment's border).
Additionally, one can end up with two O-points per watershed segment, but this is straightforwardly mitigated by the contour-finding algorithm as it only selects the contour enclosing the O-point in question. 
Even though we may sacrifice some accuracy, our methodology saves us from having to employ (relatively) computationally and memory intensive mitigation strategies and will therefore provide a significant speed-up with respect to, e.g., \citet{servidio10}.

\subsection{Harris sheet configuration} \label{meth:harris}
To validate the methodology for a well-known case, we investigate a relativistic 2D Harris sheet in resistive MHD.
The implementation is broadly based on what was prescribed for the Geospace Environmental Modeling (GEM) challenge (\citealt{birn01,birnhesse01}, also in \citealt{goedbloed10}). 
We start with a (wide) rectangular box with periodic boundary conditions on all sides and initialize two sheets of matter on top of an uniform background density that is scaled with $\rho_0$;
\begin{equation}
  \rho = \rho_0 \left[ \cosh^{-2} \left( \tfrac{y + L_y/2}{\delta} \right) + \cosh^{-2} \left( \tfrac{y - L_y/2}{\delta} \right) + f_\textrm{bg} \right], \label{eq:harris_rho} 
\end{equation}
where $L_x$, $L_y$, $f_\textrm{bg}$, and $\delta$ are the box half-size in $\hat{x}$ and $\hat{y}$, the background factor, and the layer half-thickness, respectively.
The initial values that were used for these parameters (and others) are denoted in Table \ref{tab:harris}.

\begin{table}
    \centering
    \begin{tabular}{ cccccccc }  \toprule
      \emph{Name} & $\rho_0$ & $\delta$ & $L_x$ & $L_y$ & $f_\textrm{bg}$ & \emph{Effective Res.} & \emph{AMR} \\ 
      & & & [$l$] & [$l$] & & $N_x \times N_y$ & \emph{levels} \\\midrule
      \tt{Hs} & 1 & 0.1  & 25.6 & 12.8 & 0.2 & $6144 \times 3072$  & 6 \\  
      \tt{Hb} & 1 & 0.05 & 51.2 & 12.8 & 0.2 & $24576 \times 6144$ & 5 \\ \bottomrule
    \end{tabular}
    \caption{The user-defined initial parameters for the Harris sheet simulations that include the model names (acronyms are derived from \textit{Harris-small} and \textit{Harris-big}), density scaling $\rho_0$, layer half-thickness $\delta$, background factor $f_\textrm{bg}$, and resolution (with corresponding AMR level). The total dimensions of the box are denoted by $x \in [-L_x, \: L_x]$, $y \in [-L_y, \: L_y]$. In addition to the listed parameters, there are several parameters that are constant between (all) the simulations; magnetic field scaling $B_0=1$, resistivity $\eta = 5 \cdot 10^{-5}$, and adiabatic index $\hat{\gamma} = 13/9$. }
    \label{tab:harris}
\end{table}

We assume an uniform resistivity; $\eta = 5 \cdot 10^{-5}$, and an initialized magnetic and electric field according to
\begin{align}
  &B^x = \left\{
  \begin{aligned}
    &B_0 \tanh \left( \tfrac{y - L_y/2}{\delta} \right) + B_0 \epsilon_p  & \text{for} \, & \, \, \, \, \, y > 0 \\
    -&B_0 \tanh \left( \tfrac{y - L_y/2}{\delta} \right) + B_0 \epsilon_p & \text{for} \, & \, \, \, \, \, y < 0
  \end{aligned} \, \, \, \, , 
  \right. \label{eq:harris_Bx} \\
  &B^y = B_0 \epsilon_p, \label{eq:harris_By} \\
  &B^z = 0, \\
  &E^x = E^y = E^z = 0.
\end{align}
Here, $\epsilon_p$ denotes a ($1\%$) white noise perturbation to the magnetic field that varies between $-0.01$ and $0.01$. 
This perturbation is similar to what is introduced (more naturally) for PIC simulations.
Note that we do not apply the typical guiding magnetic field perturbation that guides the initial plasmoids to the edges and creates a well-controlled reconnection region in the middle of the simulation domain \citep[as perscribed for the GEM challenge, also in][]{keppens13}.
To acquire pressure equilibrium at initialization we define the fluid pressure to be
\begin{equation}
  p = \tfrac{B_0^2}{2} \tfrac{\rho}{\rho_0}.
  \label{eq:harris_p}
\end{equation}
Additionally, we define the length and time scale as a function of system length $L = 2 \, L_x$, so that $(x,y) \in [-0.5L,0.5L] \times [-0.125L,0.125L]$ for {\tt Hb} and $(x,y) \in [-0.5L,0.5L] \times [-0.25L,0.25L]$ for {\tt Hs} with a typical time unit of $t_c=L/c$ (see Table \ref{tab:harris}).
For completeness, we note that the computational length unit is $l = 1$ with corresponding time-scale $l/c = 1$, which both reduce to unity due to the geometrical unit assumption ($G=c=1$). 
If one were interested in relating the initial layer half-thickness $\delta$ (see Table \ref{tab:harris}) to the resistivity $\eta$, then one finds that $\delta / \eta = 1000$ ($\delta / \eta = 2000$) for {\tt Hb} ({\tt Hs}).

Nevertheless, we will connect it to a more intrinsic plasma-physical timescale in our unit set in the following paragraph. 
This is typically determined by the upstream Alfv\'en velocity $v_a$, which is defined as
\begin{equation}
  v_a = \frac{B}{\sqrt{\rho h + B^2}} = \frac{\sqrt{\sigma}}{\sqrt{1 + \sigma}},
  \label{eq:harris_alfven_velocity}
\end{equation}
where $h = 1 + \hat{\gamma} p / (\hat{\gamma} - 1) \rho$ is the specific enthalpy with adiabatic index $\hat{\gamma}=13/9$ and $B = \sqrt{B^2} = \sqrt{B^iB_i}$ denotes the magnetic field strength.
Additionally, the ("hot") magnetization is defined as $\sigma=B^2 / \rho h$.
While we will primarily use the light-crossing time, it is worthwhile to connect it to the Alfv\'en and (resistive) diffusion timescales of the system, which then become $\tau_a \approx L' / v_a$ and $\tau_d \approx {L'}^2 / \eta$ with $L'$ being the current sheet's length \citep{ripperda19b}.
Figure \ref{fig:haroverview} gives an overview of the evolution of the Harris sheet (for the {\tt Hb} case).
From the magnetization ($\sigma$) panels, we find that $\sigma \sim 5$ near the sheet, which indicates an upstream Alfv\'en velocity $v_a \sim c$.
Then, one can determine the Lundquist number via $S=\tau_d / \tau_a$, but it becomes clear $\tau_d$ is very large and $\tau_a \sim L'$ which indicates that $S$ will be similarly large.

Lastly, we would like to note that all boundaries are fully periodic 
(similar to \citealt{keppens13,takamoto13,cerutti13,cerutti14a}, and some quasi-periodic works in \citealt{sironi14,petropoulou18}).
This implies that no matter is lost so that evolution eventually saturates after having formed several `monster' plasmoids that effectively act as a reservoir spanning a large part of simulation domain.
Up to a point, each sheet will evolve independently and uniquely due to the minor non-uniform perturbation to the initialized magnetic field, but when the primary plasmoids become too large the sheets are influenced by one another. 
Another approach has outflowing boundaries at the short sides of the box corresponding to the y-boundaries in our simulation \citep{loureiro12,sironi16}.
This tends to give less chaotic current sheets and allows for longer evolution times as, for periodic boundaries, the large plasmoids will eventually affect the opposing current sheet.
The periodic Harris sheet simulations are primarily meant to have another verification case for the identification algorithm, but tend to display more complex behavior than what is found for the outflowing variety, especially combined with a global magnetic field perturbation \citep[so that $\mathrm{sign}(x) \! \cdot \! u_x \gtrsim 0$;][]{loureiro12,sironi16}. 
Nevertheless, we did make sure that the magnetization was comparable to the GRMHD simulations.

\subsection{GRMHD configuration} \label{meth:grmhd}
To evolve the accretion disc surrounding the BH we utilize the Modified Kerr-Schild (MKS) coordinate system \citep[that is clearly described in][]{mckinney04,porth17}.
As the Kerr-Schild (KS) metric is well-documented \citep{misner73}, we will only comment on the modification from the standard KS coordinates ($t,r,\theta,\phi$), which is done via;
\begin{align}
    &r      = R_0 + e^s, \\
    &\theta = \vartheta + \tfrac{h}{2} \sin (2 \vartheta).
\end{align}
Here, $s$ and $\vartheta$ are the code's internally used coordinates, which can be converted to KS coordinates with the listed relations.
We will exclusively show results in KS coordinates $r$ and $\theta$.
All our GRMHD simulation use user-defined parameters $h=0.25$ and $R_0=0$, which implies that the resolution of the underlying grid will be more concentrated in the equatorial plane.

Before continuing, we would like to outline a few specifics about the $3+1$ split that is employed in \bhac. The line element is described as follows;
\begin{equation}
    ds^2 = - \alpha^2 dt^2 + \tensor{\gamma}{_i_j} ( dx^i + \beta^i dt ) ( dx^j + \beta^j dt), \label{eq:line_element}
\end{equation}
with $\alpha$, $\beta$, $\gamma$ denoting the lapse, shift, and geometric part of the metric ($g^{\mu\nu}$), where Roman characters $i,j \in \{1,2,3\}$ and Greek characters $\mu,\nu \in \{0,1,2,3\}$.
The metric determinant is then defined as $\sqrt{-g}=\alpha\sqrt{\gamma}$.
Consistent with the conventions introduced in \citet{porth17}, we denote electromagnetic quantities in the Eulerian frame with capitalized letters while lower-case letters denote quantities in the co-moving fluid (or plasma) frame.
With Eulerian frame, we imply an Eulerian observer that is moving with four-velocity $n_\mu = \{-\alpha, 0, 0, 0\}$ (or contravariantly; $n^\mu = \{1/\alpha, \beta^i/\alpha\}$).

In this work, we will only consider Magnetically Arrested Disc \citep[MAD,][]{igumenshchev03,narayan03} models which are initialized via the vector potential
\begin{equation}
  A_\phi \propto \max\left( \frac{\rho}{\rho_{\rm max}}\left(\frac{r}{r_{\rm in}}\right)^3 \sin^3 \theta \exp \left(-\frac{r}{400}\right) - 0.2, \: 0 \right) \, . \label{eq:aphi}
\end{equation}
The simulations are initialize with a torus that is in hydrodynamic equilibrium \citep{fishbone76}, except for a perturbation to the fluid pressure $p$, and is threaded by a single poloidal magnetic field loop (that is initialized via $\bf{B} = \nabla \times \bf{A}$ with ${\bf A} = (0,0,A_\phi)$).
The inner and pressure maximum radii of the torus that determine the size and available matter are set to $r_\mathrm{in}=20r_g$ and $r_\mathrm{max}=41r_g$ for a black hole spin of $a_\ast = 0.9375$.
The scaling of the vector potential is set so that $\beta = p / p_\mathrm{mag} = 100$, with $p$ being the gas pressure and $p_\mathrm{mag}$ the magnetic pressure.
Other user-defined parameters of the evaluated configurations can be found in Table \ref{tab:GRMHDtable}. 
For completeness, we will note that the less magnetized accretion scenario is known as the Standard And Normal Evolution model \citep[hereafter SANE,][]{devilliers03,narayan12,sadowski13}.

\begin{table}
    \centering
    \begin{tabular}{ ccccc }  \toprule
      \emph{Name} & \emph{Type} & $\eta$ & \emph{Effective Resolution} & \emph{AMR} \\  
      & \emph{GRMHD} & & $N_r$, $N_\theta$ & \emph{levels} \\ \midrule
        \tt{iM3} & Ideal     &  -                 & $2048 \times 2048$ & 3  \\  
        \tt{iM4} & Ideal     &  -                 & $4096 \times 4096$ & 4  \\  
        \tt{iM5} & Ideal     &  -                 & $8192 \times 8192$ & 5  \\  
        \tt{rM3} & Resistive &  $5 \cdot 10^{-5}$ & $2048 \times 2048$ & 3  \\  
        \tt{rM4} & Resistive &  $5 \cdot 10^{-5}$ & $4096 \times 4096$ & 4  \\  
        \tt{rM5} & Resistive &  $5 \cdot 10^{-5}$ & $8192 \times 8192$ & 5  \\  \bottomrule
    \end{tabular}
    \caption{The model names and corresponding resolutions of the GRMHD simulations. These simulations are all run with a dimensionless black hole spin $a_* = 0.9375$, adiabatic index $\hat{\gamma}=13/9$, and simulation domain $r \in [1.185 r_\mathrm{g}, \: 1500 r_\mathrm{g}]$, $\theta \in [0, \: \pi]$. The density floor and magnetization ceiling are set to $\rho_\textrm{min} = 10^{-4}$ and $\sigma_\textrm{max} = 10^3$, respectively. }
    \label{tab:GRMHDtable}
\end{table}

\subsection{Energetics} \label{meth:energetics}

An important objective in this work is to quantify if plasmoids are able to produce flaring events or create hot spots that would stand out with respect to the background.
Therefore, we associate the electro-magnetic, kinetic, and thermal fluid energies with their corresponding components of the stress-energy tensor $\tensor{T}{^\mu^\nu}$; 
\begin{align}
  &\epsilon_\textrm{em}  = -\tensor{T}{_{\mathrm{EM}}^{t}_{t}} = -(b^2 + e^2) (u^t u_t + \tfrac{1}{2} \tensor{g}{^t_t}) + b^t b_t + e^t e_t \label{eq:eem} \\
  & + \tfrac{u_\lambda e_\beta b_\kappa}{\sqrt{\gamma}} \left( u^t \tensor{\eta}{_t^{\lambda \beta \kappa}} + u_t \tensor{\eta}{^{t \lambda \beta \kappa}} \right), \nonumber \\
  &\epsilon_\textrm{kin} = -\tensor{T}{_{\mathrm{PAKE}}^t_t}   = - \left( u_t + 1  \right) \rho u^t \label{eq:ekin},\\
  &\epsilon_\textrm{th}  = -\tensor{T}{_{\mathrm{EN}}^t_t}     = - (\epsilon + p) u^t u_t - p  \label{eq:etherm}.
\end{align}
Here, the hereto unexplained quantities are $\epsilon$, $p$, and $\eta^{\nu \lambda \beta \kappa}$, which are the specific interal energy, the fluid pressure, and the fully antisymmetric symbol, respectively.
$\epsilon_\textrm{em}$ denotes the electro-magnetic energy density \citep{qian17}, $\epsilon_\textrm{kin}$ the kinetic energy density, and $\epsilon_\textrm{en}$ the thermal energy density \citep{mckinney12,ripperda19}.
The subscripts ``EM'', ``PAKE'' and ``EN'' correspond to the electro-magnetic, free particle, and enthalpy terms of the stress-energy tensor $T^{\mu\nu}$ \citep[primarily following][]{mckinney12}.
The free thermokinetic energy (denoted as ``MAKE'' in \citealt{mckinney12}) is the sum of $\epsilon_\mathrm{kin}$ (``PAKE'') and $\epsilon_\mathrm{th}$ (``EN'').
This is important to note because $\epsilon_\mathrm{kin}$ is predominantly negative in our GRMHD simulation, which can be interpreted from the geometric Bernoulli criterion ($u_t \leqslant -1$) corresponding to unbound matter.
The term $(u_t + 1)$ will therefore be negative (positive) when the fluid element is unbound (bound) and as $\rho u^t$ is positive we will end up with a negative $\epsilon_\mathrm{kin}$ for bound matter that is typically found within the accretion disc.
Lastly, note that the minus-sign in front of $\tensor{T}{^t_t}$ is due to the metric signature $(-,+,+,+)$ and is needed to get positive values.

Next, we define the covariant surface average (denoted by a bar, $\thickbar{\mathcal{Q}}$, over a given fluid variable) by
\begin{equation}
  \thickbar{\mathcal{Q}} = \frac{\int\sqrt{\gamma}\, \mathcal{Q} dx^1 dx^2} {S} 
  \label{eq:surface_average}
\end{equation}
with the surface $S$, in an arbitrary coordinate system, denoted as
\begin{equation}
  S = \int \sqrt{\gamma} dx^1 dx^2.
  \label{eq:surface}
\end{equation}
The $\gamma$ corresponds to the geometric part of the metric as explained in section \ref{meth:grmhd}.
Note that by surface average we imply that we take the average of a given quantity that is enclosed by a plasmoid-describing contour found by the algorithm.
All quantities are calculated in the Eulerian (or laboratory) frame.

\section{Results} \label{sec:results}

\subsection{Harris sheet}

\subsubsection{General evolution}

\begin{figure*}
  \centering
  \includegraphics[width=\textwidth]{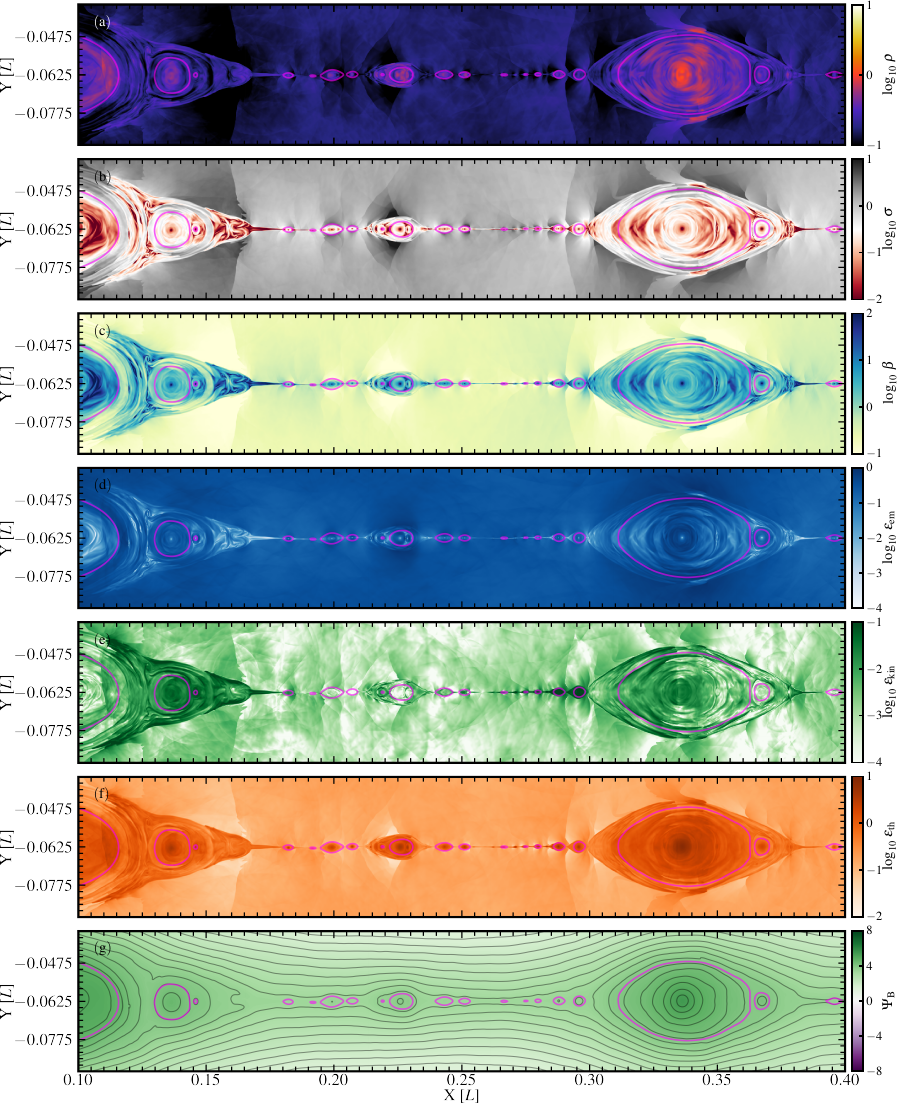}
  \vspace{-0.3cm}
  \caption
  {Representative state for the evolution of the Harris sheet for the {\tt Hb} case corresponding to $T=2.46 \, t_c$. Rows (a) till (g) show the density $\rho$, `hot' magnetization $\sigma=B^2/\rho h$, plasma $\beta = p/(B^2/2)$, electro-magnetic energy density $\epsilon_\mathrm{em}$, kinetic energy density $\epsilon_\mathrm{kin}$, thermal energy density $\epsilon_\mathrm{th}$, and magnetic flux function \psib. The \emph{purple} contours denote plasmoid detections corresponding to local maxima in the flux function (\psib), while \emph{green} contours correspond to local minima. The evolution over time is displayed in two animations; one for the zoom-in corresponding to this figure and another displaying the entire simulation domain, which can be found in the following repository; \url{https://doi.org/10.5281/zenodo.8318522}.
  }
  \label{fig:haroverview}
\end{figure*}

In Fig. \ref{fig:haroverview}, a well-developed and representative state of the {\tt Hb} case is shown.
Before this state is reached, the current sheet needs to evolve for some time before it becomes (``plasmoid-'' or ``tearing-'')unstable enough, as the sheet becomes thinner, to break up and form the first magnetic islands.
This first tearing mode creates the first plasmoids that are known as ``primary'' plasmoids \citep[see, e.g.,][]{loureiro07,uzdensky16,comisso16,petropoulou18} and have significantly different plasma characteristics than the ones that are created at later times in the secondary tearing-unstable regions of the sheets.
First, they have a higher density and, second, they possess a characteristic magnetic field profile with a lower magnetic field strength at the center than in the rings further on the outside.
Overall, this results in a lower overall magnetization, but also a relatively lower magnetic field strength in relation to the surface.
Their composition is primarily determined by the initial conditions. 
Following the initial break-up of the layer (at $\sim \!\!\! 1.56 \, t_c$ for {\tt Hb} and $\sim \!\!\! 5.27 \, t_c$ for {\tt Hs}), a continuous and steady creation of ``secondary" plasmoids has started in the reconnection layer between the primary islands that remains active till the very end of the simulation window. 
These plasmoids do probe the underlying plasma characteristics and are relatively unaffected by the initial conditions.
Two animations are attached to Fig. \ref{fig:haroverview} which show both a window correponding to the figure and the entire simulation domain over time. 

Following the formalism by \citet*{uzdensky10} (hereafter \citetalias{uzdensky10}) that when a plasmoid coalesces with a larger plasmoid, then the smaller one is considered to be part of the larger body, and is therefore no longer taken into account.
In practice, however, the small plasmoid will retain it's structure for some time depending on the size (ranging several 0.05 $t_c$) before conforming to the global structure of the primary plasmoid.
This is clearly illustrated in Fig. \ref{fig:haroverview} and accompanying animations, the coalescence of the plasmoid on the left-hand side (at $X = 0.135 \, L$ and is roughly $0.02\,L$ in width initially) takes approximately $\mathcal{O}(0.1 \, t_c)$ from the moment of impact to being fully absorbed by the primary plasmoid.
When two plasmoids of similar size coalesce, then this timescale tend to be even longer and significant perturbation is needed before one of the two loses it's structure.

Generally, it is not simple to enforce the \citetalias{uzdensky10} criterion, which is reflected by the two-step approach outlined in section \ref{meth:identification}.
Starting with secondary plasmoids, the minimum size for which we identify this population is set to $\sim \!\! 10^{-4} L$ ($0.005 \, l$), but in practice the algorithm tends to detect a plasmoid when it starts to deviate from the straight current sheet configuration (i.e., gain some width). 
Overall, we find that the secondary plasmoids are identified with a very high fidelity.
The primary plasmoids are typically much harder to identify as they are the end point of the inverse cascade (or plasmoid coalescence) and, therefore, act as highly turbulent plasma reservoirs that will never relax as smaller plasmoids keep colliding and merging into it.
These continuous perturbations also give rise to some magnetic reconnection within the primary plasmoid structure.
As described in section \ref{meth:identification}, we need to apply an aggressive blurring kernel to identify the global primary plasmoid structure, but we still want to pick up on the distinct plasmoid structure if they have not fully merged.
This implies that two plasmoids that have a similar magnetic flux signature (an example is seen at $X = 0.34L$) are still picked up as two separate entities even though one can argue that they are actually part of one global body, especially when following the \citetalias{uzdensky10} criterion.
At the interface of these two plasmoids one often finds new plasmoids forming.
Naturally, all previously mentioned points become less pronounced at lower resolutions as one is resolving the current sheets less well which results in less formed plasmoids and less fine-structure.

The end of the evaluated window (at $4.1 \, t_c$ for {\tt Hb} and $8.79 \, t_c$ for {\tt Hs}) is determined by the amount of interference the current sheets have on one another.
Beyond these times, the few primary plasmoids become of sufficient size that they start to incorporate the opposing current sheet.
This brings about an interesting new turbulent mode that is similar to the ABC structure described in \citet{lyutikov16}. 
Magnetic reconnection is then no longer confined to the current sheets but occurring at interfaces between the primary plasmoids that now have lost their elliptical shape and have become more hexagonal in shape.
The simulation has a closer resembles to a turbulent box simulation than to the initial double current sheet configuration.
This is beyond the scope of this work and therefore we chose the evaluated time windows to correspond to a clear current sheet structure. 

\subsubsection{Plasmoid statistics} \label{res:harris_plasmoid_timeseries}

\begin{figure*}
  \centering
  \includegraphics[width=0.49\textwidth]{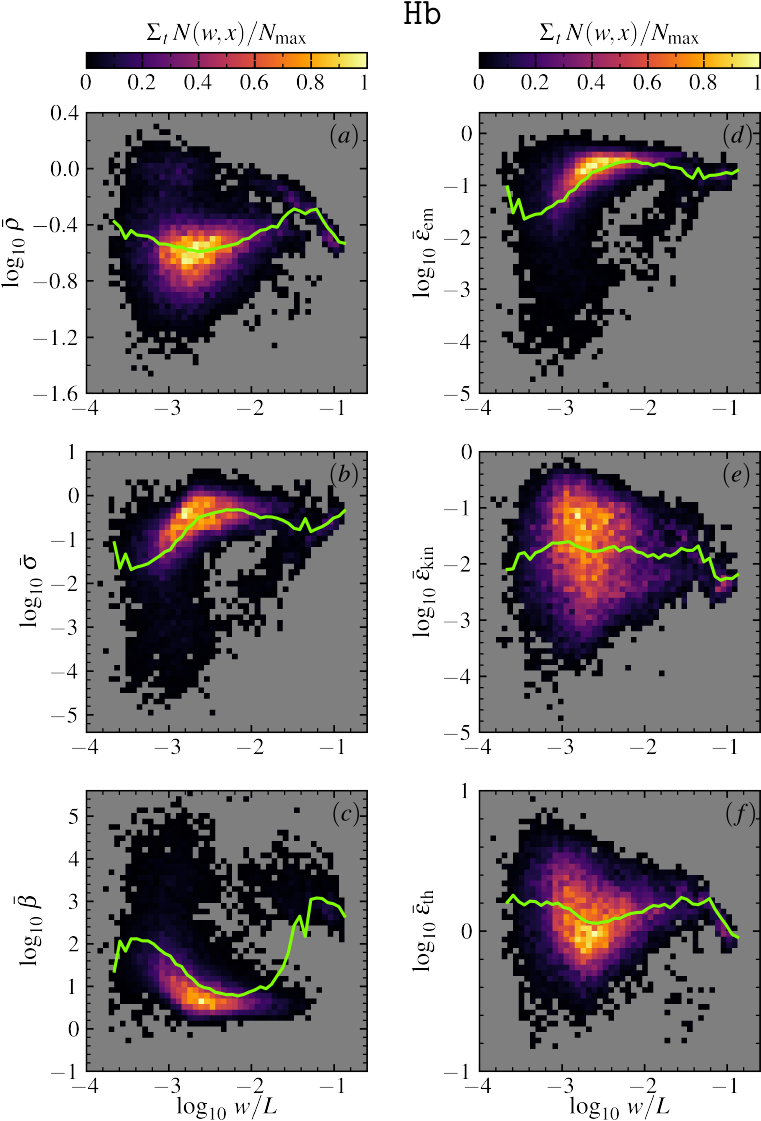}
  \hfill
  \includegraphics[width=0.49\textwidth]{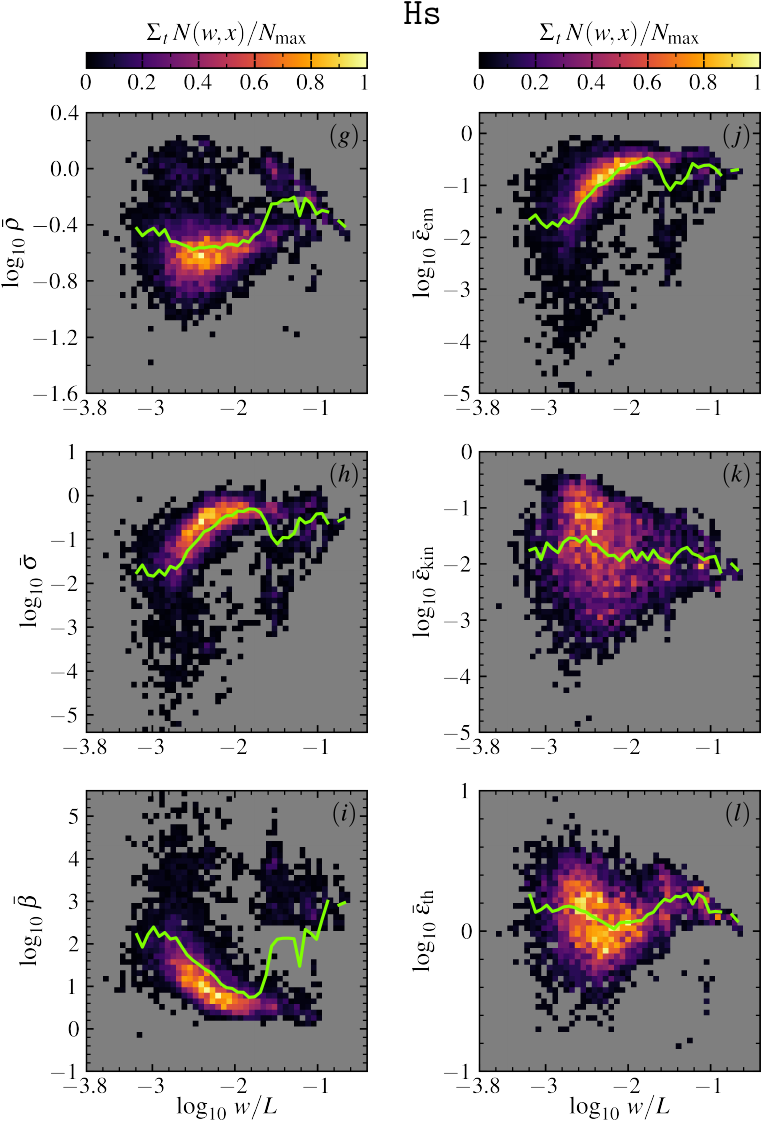}
  \caption{Two dimensional distributions $N(w,x)$ of the plasma quantities $x \in \{\bar{\rho}, \bar{\sigma}, \bar{\beta}, \bar{\epsilon}_\mathrm{em}, \bar{\epsilon}_\mathrm{kin}, \bar{\epsilon}_\mathrm{th}\}$ as a function of plasmoid half-width (with $L=2L_x$, as per Table \ref{tab:harris}) for both the {\tt Hb} (left panels) and {\tt Hs} (right panels) cases.
  We stack the distributions as a function of time and divide by the maximum.
  The green line denotes the mean per width bin that has more than ten counts total.  
  }
  \label{fig:harris_twoD_hist_Hb_Hs}
\end{figure*}

Figure \ref{fig:harris_twoD_hist_Hb_Hs} displays two-dimensional histograms with various plasmoid quantities as a function of width for both Harris sheet ({\tt Hs} and {\tt Hb}) cases.
First, we would like to point out that the distributions show the same general trends.
Starting with the surface-averaged density ($\bar{\rho}$) panels, one finds a main triangular distribution that spans $-1.25 < \log_{10}\,\bar{\rho} < -0.25$.
In addition to the main distribution, there is a secondary channel corresponding to  $-0.25 < \log_{10}\,\bar{\rho} < 0.25$ that corresponds to the densest plasmoids which also seem to occur over the entire width range. 
These plasmoids are, to summarize, in part due to minor misclassifications and due to the simulation conditions quickly after break-up of the initial layer. For the former, we find plasmoids that often correspond to fluctuation in the magnetic flux function within a large primary plasmoid. This population corresponds with a small plasmoid half-width. For the latter scenario, there are a number of high density reservoirs of matter that will eventually contract into the primary plasmoid population and will generally correspond to a large plasmoid half-width.

Returning to the ``true'' plasmoid population, spanned by $-1.25 < \log_{10}\,\bar{\rho} < -0.25$, we find that the smallest detected plasmoids have a half-width $w \approx 2 \cdot 10^{-4} \, L$ for both the {\tt Hb} and {\tt Hs} cases.
This lower limit is partially set by an identification requirement that either the width or height of the contour spans at least 5 cells (which equates to a minimal width or height of $\Delta x \approx 0.02 \, l$) for the evaluated data.
For the surface-averaged magnetization ($\bar{\sigma}$), we find that the main population spans $-2.5 < \log_{10}\,\bar{\sigma} < 0.5$.
As is also seen in Fig. \ref{fig:haroverview}, the secondary plasmoids have a remarkably similar $\sigma$ profile with the outer shells being more magnetized than the interior \citep[similar to findings in][]{petropoulou18}.
Nevertheless, we do find a trend where the $\bar{\sigma}$ rises with half-width, up to $w \approx 6.3 \cdot 10^{-3} L$.
For $\log_{10} w/L > -2$, the $\bar{\sigma}$ mean plateaus and even seems to decrease slightly for the largest plasmoids.
After the growth phase (in $\log_{10} w/L \lesssim -2$), it seems that the increase in density and magnetic field strength is roughly matched.
Lastly, for $\bar{\beta}$, we find a similar but inverse trends to what we described for $\bar{\sigma}$.
The part of the distribution with the largest plasmoids ($w \sim 0.1 \, L$) seems to deviate significantly from the main population and possesses a relatively high $\bar{\beta} \gtrsim 10^3$.
This happens because at the center of the plasmoid the magnetic field strength becomes very small due to the circular configuration.
This generates some very high $\beta$ values that in turn affects the surface-averaged quantity ($\bar{\beta}$).

For the energies ($\bar{\epsilon}_\mathrm{em}$, $\bar{\epsilon}_\mathrm{kin}$, and $\bar{\epsilon}_\mathrm{th}$), we find that the thermal energy ($\epsilon_\mathrm{th}$) is the leading term in the total energy budget of the plasmoids with a mean (denoted by the green line) that remains fairly constant ($0.0 < \log_{10}\,\bar{\epsilon}_\mathrm{th} < 0.25$) as a function of half-width ($w$).
At smallest $w$, it appears the second term is the electro-magnetic energy (at $\bar{\epsilon}_\mathrm{em} \approx 10^{-1.5}$) that steadily becomes more significant for increasing width.
As the kinetic energy ($\epsilon_\mathrm{kin}$) is closely tied to the velocity of the plasmoid, we find that it can actually become a competing term for the electro-magnetic energy, especially in the active reconnection regions and merging (or colliding) plasmoids (see Fig. \ref{fig:haroverview}).
The distribution of $\bar{\epsilon}_\mathrm{th}$ and $\bar{\epsilon}_\mathrm{kin}$ are wide indicating significant variance, while $\bar{\epsilon}_\mathrm{em}$ closely follows the distribution of $\bar{\sigma}$ and seems to show a more consistent trend.
This trend is explained by secondary plasmoids becoming more magnetized with time, until they grow up to a size of $w \sim 0.01 L$, after which they generally encounter a primary plasmoid and are absorbed after which the growth in magnetization ($\bar{\epsilon}_\mathrm{em}$) stagnates.
The high variance in $\bar{\epsilon}_\mathrm{kin}$ is explained by the fact that acceleration of plasmoids only happens in very localized regions -- predominantly in active reconnection regions and just before plasmoids coalesce. 
As soon as the secondary are absorbed by the primary plasmoids, $\bar{\epsilon}_\mathrm{th}$ will be the leading term by a significant factor.
Even though $\bar{\epsilon}_\mathrm{th}$ is still most dominant in the secondary plasmoid, both $\bar{\epsilon}_\mathrm{kin}$ and, especially, $\bar{\epsilon}_\mathrm{em}$ can become close in significance.

Lastly, we would like to briefly comment on the differences between the two cases; {\tt Hb} and {\tt Hs}.
So far, we have mainly talked about the {\tt Hb}, in the left-most panels of Fig. \ref{fig:harris_twoD_hist_Hb_Hs}.
Nevertheless, we find that all findings are also applicable to {\tt Hs}.
The description of both simulations is outlined in Table \ref{tab:harris}, where we find that the main differences lie in the initial layer (half-)thickness ($\delta$) that is twice as wide and resolution that is lower by a factor two.
This also explains why the evolution starts later for {\tt Hs}; it takes longer for the perturbations to create a sufficiently thin current sheet to activate the tearing instability.
Additionally, simulation box length (in $\hat{x}$, long side) is halved and it contains more matter due to the thicker initial layers when compared to the {\tt Hb} case.
As these are only minor differences, we find that the evolution is similar, which is also reflected by the results here, except that the primary plasmoids seem to span a greater part of the simulation domain for {\tt Hs}.
To gain insight into the dependence of the plasmoid dynamics on starting conditions, a more detailed study is needed, but that lies beyond the scope of this work. 

\subsubsection{Plasmoid distribution functions} \label{res:harris_plasmoid_distribution_function}

\begin{figure*}
  \centering
  \includegraphics[width=0.49\textwidth]{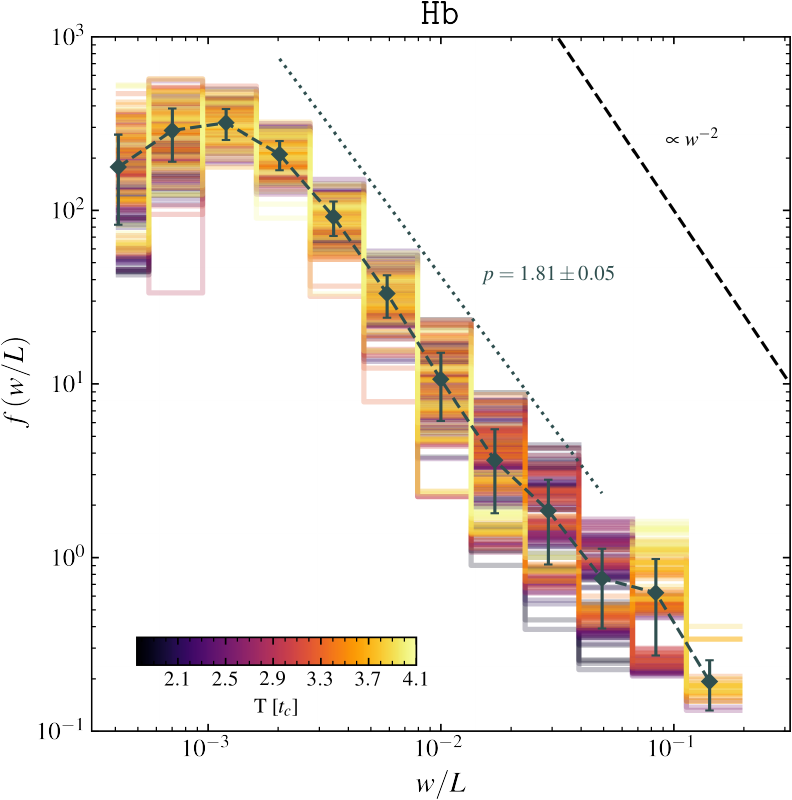}
  \hfill
  \includegraphics[width=0.49\textwidth]{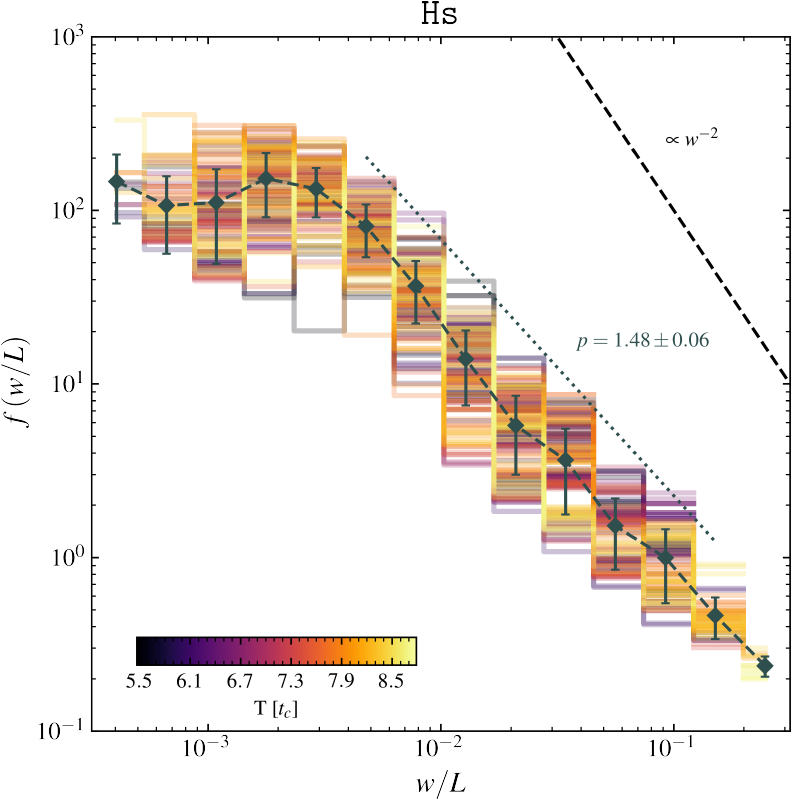}
  \caption{Probability density function $f (w / L)$ of plasmoid half-width $w$ (along $\hat{x}$) that is scaled according to the total simulation box width ($L = 2L_x$) for both the Harris sheet cases; {\tt Hb} and {\tt Hs}.
    From scaling arguments, proposed in \citep{uzdensky10,loureiro12}, it has been shown that the distribution is expected to scale $\propto w^{-2}$. The various probability density function profiles are colored according the time at which they occur over a range of $T \in [1.76,4.10] \, t_c$ for {\tt Hb} and $T \in [5.47,8.79] \, t_c$ for {\tt Hs}. The mean density profile and one-sigma error over time are denoted by the dashed dark grey line. The power law slope is determined via $p = - \, \mathrm{d} \log f / \, \mathrm{d} \log (w/L)$.
  }
  \label{fig:harris_pdf_width}
\end{figure*}

\begin{figure*}
  \centering
  \includegraphics[width=0.49\textwidth]{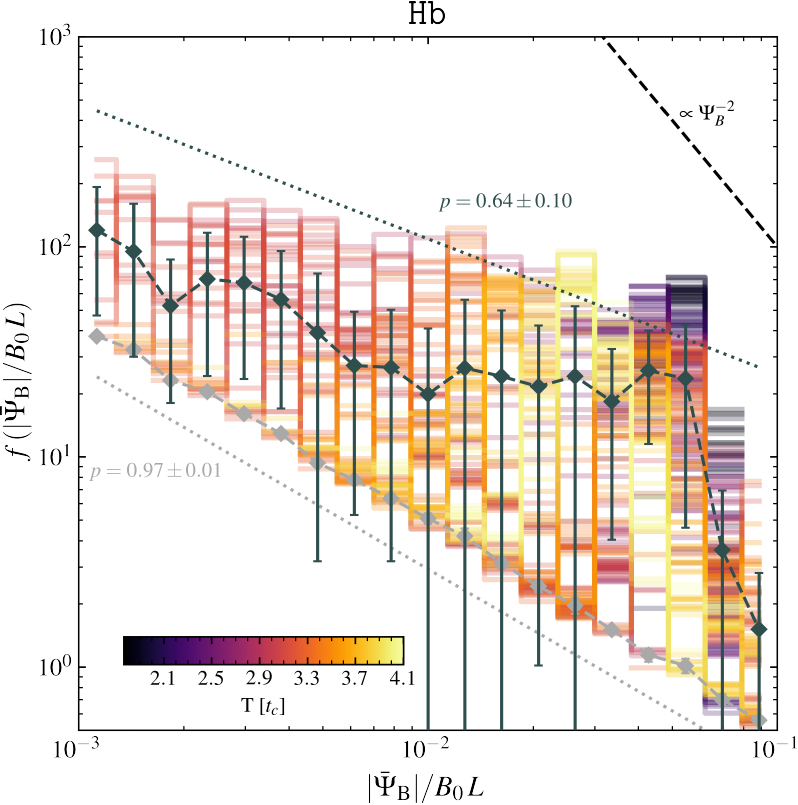}
  \hfill
  \includegraphics[width=0.49\textwidth]{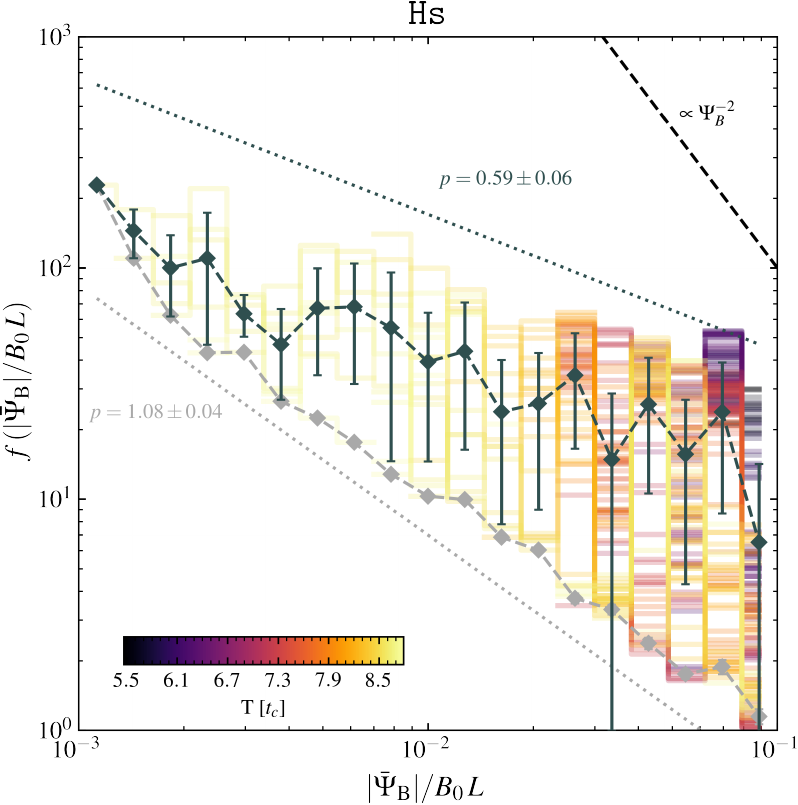}
  \caption{Probability density function $f (| \bar{\Psi}_\mathrm{B} | / B_0 L)$ of plasmoid half-width $| \bar{\Psi}_\mathrm{B} | / B_0 L$ that is scaled according to the total simulation box width ($L = 2L_x$) and initial magnetic field strength ($B_0$) for both the Harris sheet cases; {\tt Hb} and {\tt Hs}. The light grey dashed lines denote the mean of the lowest ($1\%$) values per bin with a corresponding linear fit shown in the same color. For the rest, the description of Fig. \ref{fig:harris_pdf_width} also applies here.}
  \label{fig:harris_pdf_psib}
\end{figure*}

Figures \ref{fig:harris_pdf_width} and \ref{fig:harris_pdf_psib} display the probability density function ($f$) of plasmoid half-width and the absolute surface averaged magnetic flux function ($|\bar{\Psi}_\textrm{B}|$), respectively.
A distribution is calculated at each time starting at the beginning of the evalutated window at $T = 1.76 \, t_c$ for {\tt Hb} ($T = 5.47 \, t_c$ for {\tt Hs}) in dark blue up to $T = 4.1 \, t_c$ ($T = 8.79 \, t_c$) in bright yellow.
Starting with Fig. \ref{fig:harris_pdf_width}, we find there is reasonable variation between the probability density function over time, but a consistent image emerges as well.
Generally speaking, at the smallest plasmoid half-widths (up to $w / L \approx 10^{-3}$), we find a plateau followed by a steady decrease in occurrence frequency as the plasmoids become larger, up to the largest plasmoids that span a tenth of the simulation domain ($w \sim 0.1 L$).
Mainly via the slope $p = - \mathrm{d} \log f / \mathrm{d} \log (w / L)$, we will be able to quantity the growth rate of plasmoids in the system.

The scaling laws have been studies in detail in the past \citep{uzdensky10,loureiro12,huang12,sironi16}.
The density function of plasmoid width was predicted and verified to scale according to $f(w) \sim w^{-2}$ \citep{uzdensky10,loureiro12}, while for magnetic flux both $f(\Psi_\mathrm{B}) \sim \Psi_\mathrm{B}^{-2}$ (following the same works) or $f(\Psi_\mathrm{B}) \sim \Psi_\mathrm{B}^{-1}$ \citep{huang12} were established.
The main difference between scaling found by \citet{uzdensky10} and \citet{huang12} lies in how they treat the relative velocity between plasmoids.
While \citet{uzdensky10} assumed it to be $\sim \!\! v_a$, \citet{huang12} evaluates a size-dependent relative velocity \citep[see also][]{sironi16}.
As our simulations have no guiding magnetic field perturbation (or outflowing boundaries), relative velocities between plasmoids are stochastically determined and relatively low, so we expect a greater similarity with \citet{huang12}.
Overall, we find that \psib and $w$ do not scale with the same $p$, which is contradictory with earlier works \citep{loureiro12,sironi16}.
However, there are clear explanations for this perceived discrepancy that will be outlined in the next paragraphs.

For half-width, we find $p = 1.81 \pm 0.05$ for the {\tt Hb} case and $p = 1.48 \pm 0.06$ for the {\tt Hs} case.
Overall, for $w$, we find that that we recover a scaling that is close to $f(w) \sim w^{-2}$ corresponding to $p=2$.
For the mean trend in magnetic flux (in dark grey), we find $p = 0.64 \pm 0.10$ for the {\tt Hb} case and $p = 0.59 \pm 0.06$ for the {\tt Hs} case.
However, the trend described by the smallest values per bin (in light grey) is $p \approx 1$, which indicates agreement with \citet{huang12}.
The evolution of the distributions is characterized by a relative over-representation of large plasmoids, with $|\Psi_\mathrm{B}/B_0 L| \in [5 \cdot 10^{-2}, 10^{-1}]$, that expands itself both to the left (lower $|\Psi_\mathrm{B}|$, smaller plasmoids) and right (higher $|\Psi_\mathrm{B}|$, larger plasmoids) over time.
The smallest plasmoids have the lowest magnetic fluxes (as is also verified in Fig. \ref{fig:haroverview}) and the largest plasmoid will increase in $|\Psi_\mathrm{B}|$ over time.
This evolution also creates the sizable one-sigma error (visually made worse by the log-scale) as the density function evolves significantly over time.
So, in short, the magnetic flux distributions evolve with $p=1$ over time (especially for a low $|\Psi_\mathrm{B}|$), but this relation is affected by a high $|\Psi_\mathrm{B}|$ population (that is present from the start). 
This population is there because of the periodic boundary conditions and would not be over-represented when utilizing outflowing boundary conditions, which was done by the comparative studies.

It is important to note that our simulation setup differs substantially on at least two fronts from the previously mentioned scaling law studies, namely that it is relatively unperturbed and that it has no outflowing boundaries.
With unperturbed, we mean that there is no guiding magnetic field perturbation present, which recreates a clean reconnection layer in the middle of the box and guides the primary plasmoids to the edge of the simulation domain (also discussed in detail in section \ref{meth:harris}).
In practice, this implies that; (i) coalescence of plasmoids is a relatively more prominent growth channel in our simulations and (ii) large plasmoids could disproportionately affect the distribution.
The latter point is two-fold; as the primary plasmoids become larger they effectively shrink the domain where the (secondary) current sheets can form and they will eventually start interfering with the opposing current sheet.
Especially for the {\tt Hs} case, these points are influential, which is also accentuated by the larger deviations.
All these effects are likely to play a role in explaining the differences in scaling found in this work with respect to previous works.
Additionally, the informed (but arbitrary) choice regarding which bins to include for the fit combined with the imperfect sampling of the distribution by the bins also introduces a $\mathcal{O}(5\%)$ error on the values of $p$.
Although, even despite the differences in simulation configuration (and the numerical uncertainties), we still reach a remarkable consistency with previous studies that employed more idealized configurations for finding plasmoid scaling.

\subsection{GRMHD}

\subsubsection{General evolution} \label{res:grmhd_general_evolution}

\begin{figure*}
  \centering
  \includegraphics[width=\textwidth]{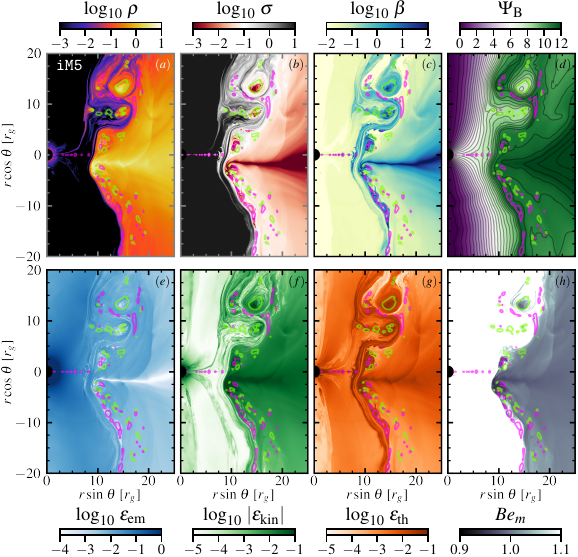}
  \caption{Overview of the {\tt iM5} simulation at $T = 3840 \, r_\mathrm{g}/c$. 
  Here, we are in the middle of a flux eruption event with pushed back the accretion disc.
  The \emph{purple} contours corresponds to local maxima and the \emph{green} contours correspond to local maxima in the magnetic flux function ($\Psi_\mathrm{B}$).
  In panels $(a)$-$(d)$, one finds the density ($\rho$), the magnetization ($\sigma$), the ideal to magnetic pressure ratio ($\beta$), and the magnetic flux function (\psib). In the panels $(e)$-$(h)$, we find the electro-magnetic energy ($\epsilon_\mathrm{em}$), kinetic energy ($\epsilon_\mathrm{kin}$), thermal energy ($\epsilon_\mathrm{th}$), and the magnetic Bernoulli factor ($Be_m = - (h + \sigma/2)u_t$). The corresponding animations can be found in the following repository; \url{https://doi.org/10.5281/zenodo.8318522}. 
  }
  \label{fig:grmhd_overview_iM5}
\end{figure*}

\begin{figure*}
  \centering
  \includegraphics[width=\textwidth]{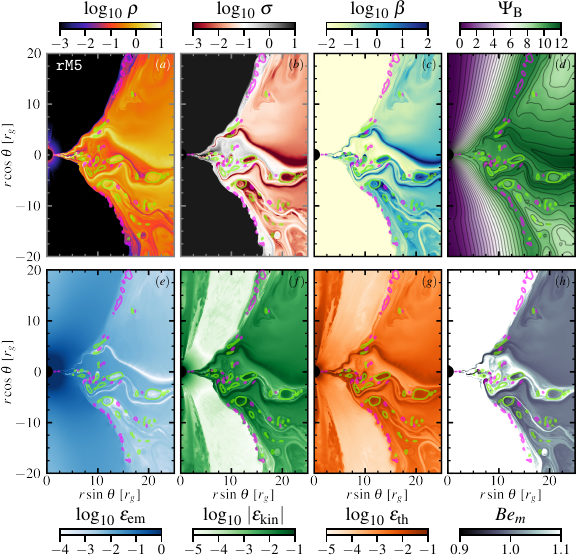}
  \caption{Overview of the {\tt rM5} simulation at $T = 3500 \, r_g/c$.
  Here, we find an accretion state that is standard for MAD simulations with a turbulent but fairly steady flow.
  The rest of the description is analogous to Fig. \ref{fig:grmhd_overview_iM5}.
  The corresponding animations can be found in the following repository; \url{https://doi.org/10.5281/zenodo.8318522}.
  }
  \label{fig:grmhd_overview_rM5}
\end{figure*}

Figures \ref{fig:grmhd_overview_iM5} and \ref{fig:grmhd_overview_rM5}  display the typical structure of the (two-dimensional) MAD \citep[][]{tchekhovskoy11,mckinney12} simulations.
After having evolved sufficiently, they will saturate in magnetic flux that penetrates the event horizon (see section \ref{res:grmhd_plasmoid_and_horizon_timeseries}).
Following such a saturation event, the accretion flow is completely halted in axisymmetric (2.5D) simulation, while in 3D a so-called ``flux tube'' forms \citep{dexter20,porth21}.
There, instead of halting the accretion flow completely, a localized less dense, more magnetized cavity moves outward from the black hole.
These outbursts occur semi-periodically and seems to be even more prevalent in the relatively more confining 2D simulations.

Another feature is that the Magneto-Rotational Instability \citep[MRI,][]{balbus94}, responsible for angular momentum transport, is suppressed as the main magnetic field component component is strongly poloidal in MAD simulations \citep{porth21}.
The MRI does play a role in the early developing phase of the simulation, when it is less magnetized, but then one of the leading causes of turbulence (close to the BH) is the Rayleigh-Taylor Instability (RTI) that causes incursions into the disc structure \citep[][and references therein]{marshall18}.
The Kelvin-Helmholtz Instability (KHI) becomes important in regions with strong shear flows as are the conditions at the jet-disc interface and are characterized by swirl-like vortices \citep[see, e.g.,][]{begelmanreview84,hillier19}.
All these instabilities are perturbatative channels that are able to set off magnetic reconnection in the accretion disc.
Therefore, we find a much more turbulent environment than for the Harris current sheet for which reconnection is only determined by the tearing instability  \citep{ripperda17a} that is triggered in a relatively controlled scenario.

As we are mainly interested in the plasmoids' ability to produce flares, which are known to originate close to the central black hole, we apply our algorithm only within the inner $25$ \rg.
In Figs. \ref{fig:grmhd_overview_iM5} and \ref{fig:grmhd_overview_rM5}, we find the plasma quantities and energies (similar to Fig. \ref{fig:haroverview}) for the {\tt iM5} and {\tt rM5} cases.
The magenta and green colors denote found plasmoid corresponding to local maximum and local minimum in the magnetic flux function, respectively.
Both figures show typical phases of MAD evolution that happen in all the GRMHD simulations in this work.
The panels ($a-h$) of Fig. \ref{fig:grmhd_overview_iM5} correspond to a flux eruption where we find the accretion flow is entirely halted.
The panels ($a-h$) of Fig. \ref{fig:grmhd_overview_rM5} show a fairly generic accretion state with the turbulent accretion flow extending up to the horizon.
Even though the density is low near the BH, one does find a reconnection layer along the equatorial plane (denoted by the magenta contours).
These plasmoids are the collisional (non-pair-production plasma) equivalent to what has been seen in GRPIC simulation of diffuse collisionless magnetospheres around BHs \citep{crinquand21,bransgrove21}.

The overall structure and location of the plasmoid chains indicate that at the disc-jet boundary one finds plasmoids that correspond to local maxima (magenta) while when plasmoids occur within the disc they correspond to local minima (green).
The magenta contours seem to have a lower density ($\rho$) and higher magnetization ($\sigma$) than the ones in the disc.
They also seem to be smaller when compared to the green contours.
Their location and smaller size indicates that they are likely created by the shear-induced KHI.
The purple contours also tend to leave the identification domain ($r \leqslant 25$ \rg) on short timescales (5--10 \rgc) as they rapidly move outwards with turbulent jet-disc layer (also referred to as jet sheath).
The green contours are tied to the bulk motion of the disc's fluid giving them more time and matter to interact with which explains their larger size.
The energy and plasma parameter distributions will be explained in more detail in the next section (\ref{res:grmhd_plasmoid_statistics}).
However, before we continue, we would like to point out that the values (visible in the $\rho$, $|\epsilon_\mathrm{kin}|$, and $\epsilon_\mathrm{th}$ maps) near the vertical axis ($x=0$ \rg) are due to floor violations, which happen sufficiently far from our areas of interest and will therefore not interfere with the analysis.

\subsubsection{Plasmoid statistics} \label{res:grmhd_plasmoid_statistics}

\begin{figure*}
  \centering
  \includegraphics[width=0.49\textwidth]{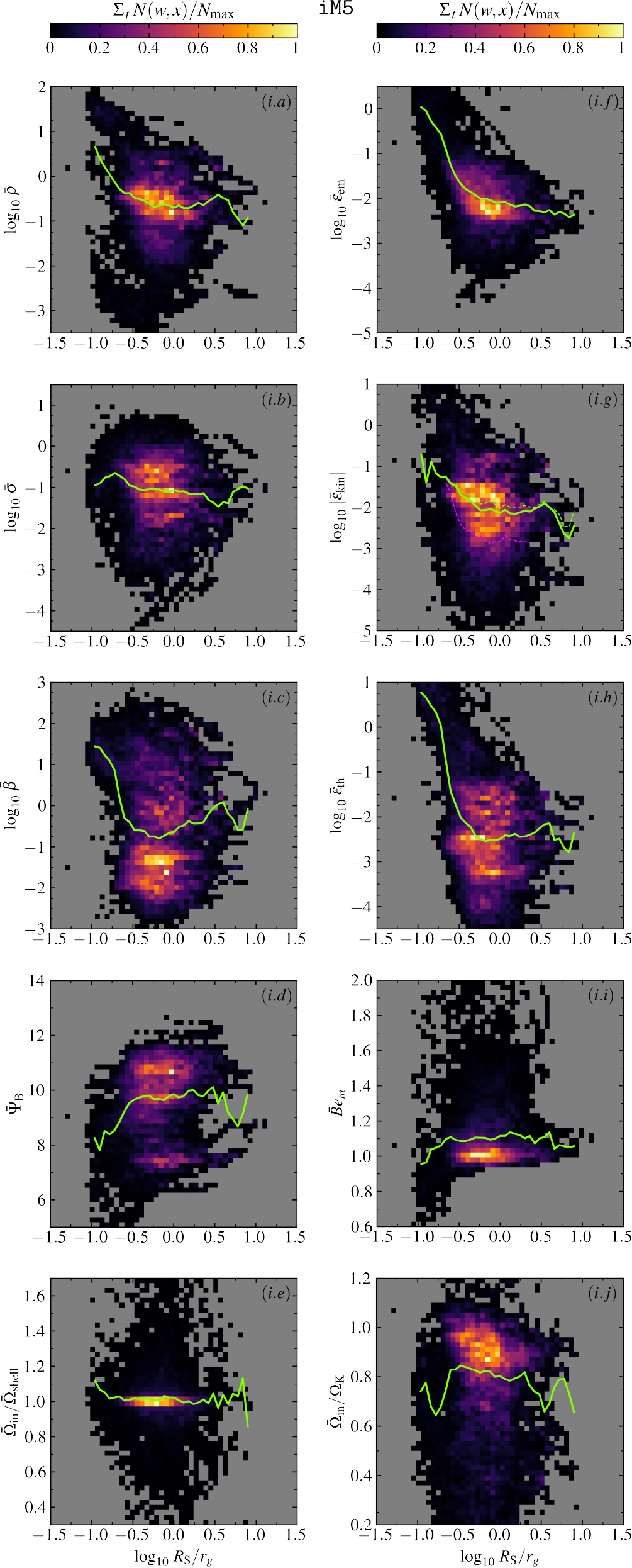}
  \hfill
  \includegraphics[width=0.49\textwidth]{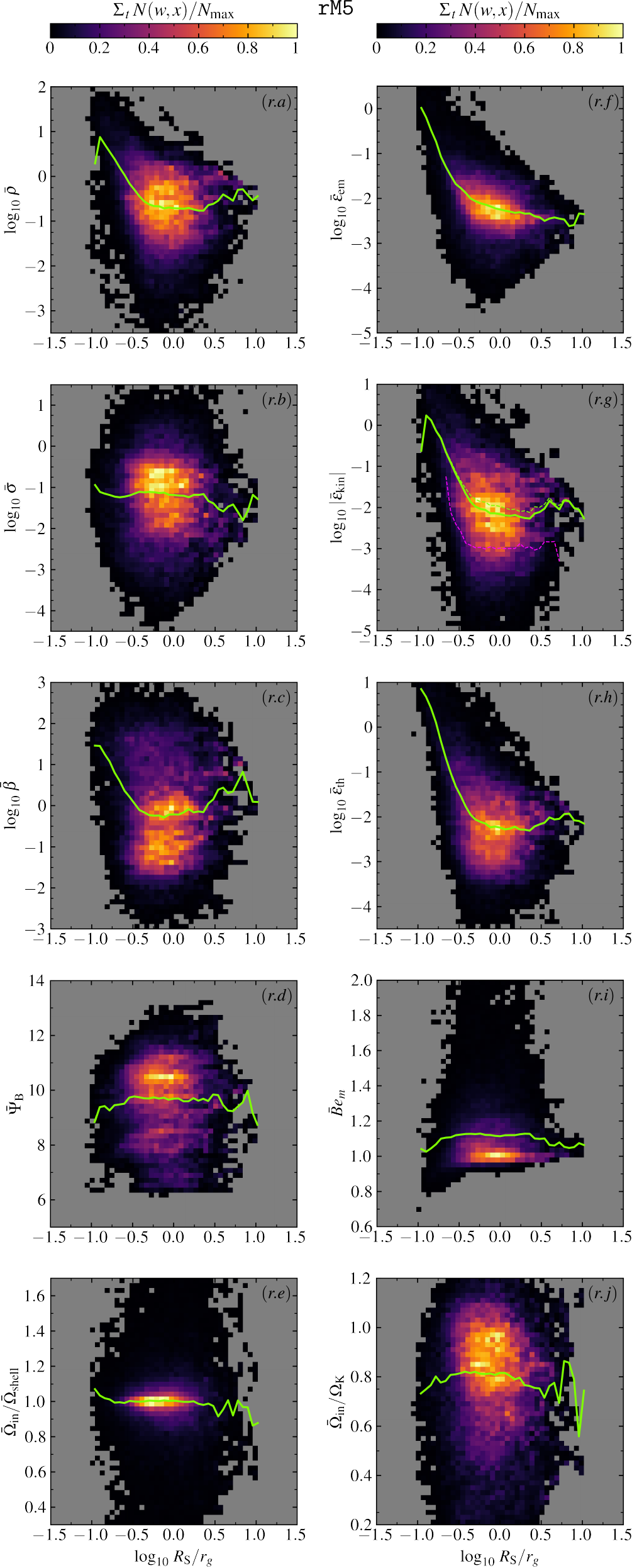}
  \caption{Two dimensional distributions $N(w,x)$ of the plasma quantities $x \in \{\bar{\rho}, \bar{\sigma}, \bar{\beta}, \bar{\Psi}_\mathrm{B}, \bar{\Omega}_\mathrm{in}/\bar{\Omega}_\mathrm{shell}, \bar{\epsilon}_\mathrm{em}, \bar{\epsilon}_\mathrm{kin}, \bar{\epsilon}_\mathrm{th}, \bar{B}e_m, \bar{\Omega}_\mathrm{in}/\Omega_\mathrm{K} \}$ as a function of ``circular'' radius $R_\mathrm{S} = \sqrt{S/\pi}$ of the plasmoid for both the {\tt iM5} and {\tt rM5} cases.
  While the other parameters have been outlined before, the magnetic Bernoulli factor is defined as $\bar{B}e_m = -(h + \sigma/2)u_t$ and the orbital velocity $\Omega = u^\phi / u^t$ with the surface-averaged quantity inside the plasmoid being denoted as $\bar{\Omega}_\textrm{in}$.
  We stack the distributions as a function of time from 3000 \rgc to 4000 \rgc with a 1 \rgc cadence and divide by the maximum.
  The green line denotes the mean per width bin that has more than 20 counts total.
  The \textit{left} panels denotes {\tt iM5} case, while the \textit{right} panels denotes the {\tt rM5} case.
  }
  \label{fig:grmhd_M5_twoD_width}
\end{figure*}

Figure \ref{fig:grmhd_M5_twoD_width} shows two-dimensional histograms with various plasmoid quantities as a function of width for the two GRMHD simulations ({\tt iM5} and {\tt rM5}).
Before we comment on the general findings from the histograms, we would like to point out that we find a significantly lower plasmoid count for the {\tt iM5} case when compared to the the {\tt rM5} case.
The ideal distributions are therefore more sparsely sampled.
We will address this point in more detail in section \ref{res:grmhd_plasmoid_and_horizon_timeseries}.
Overall, however, we do find that the distributions of {\tt iM5} and {\tt rM5} are consistent with one another, except for the aforementioned difference in occurrence rate. Before we start describing the distributions, we would like to note that we can no longer use the Euclidean width for the GRMHD cases, as this does not inherently take into account the spacetime curvature. Therefore, we have chosen to display the distribution as a function of ``circular'' radius $R_\mathrm{S} = \sqrt{S / \pi}$ as the surface $S$ calculation is taking into account the curvature.
As plasmoids are generally elliptical, we loose information about the plasmoid shape as the ratio between width and length is no longer defined.

Starting with the distribution of $\bar{\rho}$ (panels $l.a$ \& $r.a$), we find that the surface-averaged density is highest for the smallest plasmoids at $\log_{10} \, \bar{\rho} \approx 0.75$ and then plateaus at $\log_{10} \, \bar{\rho} \approx -0.5$ from $-0.5 < \log_{10} R_\mathrm{S} < 1.0$.
For $\bar{\sigma}$ ($l.b$ \& $r.b$), we find a roughly constant mean value of $\log_{10} \, \bar{\sigma} \approx -1$, but a wide spread in values is also present. 
For $\bar{\beta}$ ($l.c$ \& $r.c$), one finds a very elongated distributions centered around a mean of roughly $\log_{10} \, \bar{\beta} \approx 0.5$ which has a complicated origin.
This behavior is largely explained by the `green' (local minima in $\Psi_\mathrm{B}$) and `purple' (local maxima in $\Psi_\mathrm{B}$) plasmoid populations.
For the purple contours, we find the origin of the elongated $\bar{\beta}$ distribution as the plasmoids detected in the jet sheath correspond to a distribution centered on a relatively low $\log_{10} \bar{\beta} \approx -1$.
The typical distribution of $\bar{\beta}$ is fairly uniformly distributed with $-2 \lesssim \log_{10} \bar{\beta} \lesssim 2$ centered on the mean value of $\log_{10} \bar{\beta} \approx 0-0.5$.
Similar arguments apply to the distributions for $\bar{\rho}$ and $\bar{\sigma}$ where they both near identical means but a larger variance is present of the purple contours.
For the green contours, we find more uniform and compact distributions overall that are located around the means for the entire (both green and purple) distribution as shown in Fig. \ref{fig:grmhd_M5_twoD_width}.

For the energies ($\epsilon_\mathrm{em}$, $|\epsilon_\mathrm{kin}|$, and $\epsilon_\mathrm{th}$), there are only minor difference between the green and purple distributions, so we will just discuss the combined distributions for the energies in panels ($l.f$-$l.h$) and ($r.f$-$r.h$).
Interestingly, the mean for all energy distributions describe an almost identical path -- starting at $\log_{10} \bar{\epsilon} \approx 0$ to ending at $\log_{10} \bar{\epsilon} \approx -2$ for increasing $R_\mathrm{S}$.
After a rapid decline up to $\log_{10} R_\mathrm{S} \approx -0.5$, we find that the $\log_{10} \bar{\epsilon}$ means plateau, especially for $\bar{\epsilon}_\mathrm{kin}$ and $\bar{\epsilon}_\mathrm{th}$.
Additionally, the distributions indicate that the various surface-averaged energies are of similar strength.
Nevertheless, $\bar{\epsilon}_\mathrm{em}$ does stand out with respect to the other energies as it has a more compact distribution with a clear, gradually declining trend.
Generally speaking, we find that all energy densities are of similar strength independent of the plasmoid size.
Continuing with $\bar{\epsilon}_\mathrm{kin}$, we would like to note that $\bar{\epsilon}_\mathrm{kin}$ is negative, except in the jet-sheath where $\epsilon_\mathrm{kin} \sim \mathcal{O}(1)$.  
This is explained in detail in section \ref{meth:energetics}.
Here, we will look at the absolute value $|\bar{\epsilon}_\mathrm{kin}|$ ($l.g$ \& $r.g$).
The dashed purple and green lines in these panels correspond to the means of the distributions containing only the positive or negative values of $\bar{\epsilon}_\mathrm{kin}$, respectively.
So, it becomes clear that the vast majority of plasmoids has a negative $\bar{\epsilon}_\mathrm{kin}$ values as the global mean (in solid green) lies close to the dashed green line.
Lastly, for $\bar{\epsilon}_\mathrm{th}$, we find similar behavior as for the other energies combined with a relatively more considerable contribution at the lowest plasmoid sizes.

We define the magnetic Bernoulli factor as $Be_m = -(h+\sigma/2)u_t$, which incorporates the contribution of the magnetic pressure ($\sigma/2$) and therefore deviates slightly from the standard relativistic Bernoulli $Be = -hu_t$ \citep{rezzolla13}.
The Bernoulli criterion states that the fluid is unbound when $Be_m > 1$.
Note that we have taken the liberty to incorporate a minus sign within the Bernoulli factor.
Returning to the distributions in panels ($l.i$ \& $r.i$), we find the majority of surface-averaged plasmoids is unbound as they pass the criterion, but there is still a significant number that lies under and close to the critical value of $\bar{B}e_m = 1$ and are therefore bound. 
The mean of the function does however indicate $\bar{B}e_m \approx 1$ with a small number going up to relatively high values of $\bar{B}e_m \approx 2$. 
In the panels next to $\bar{B}e_m$, we find the distributions of $\bar{\Psi}_\mathrm{B}$ which seem elongated and somewhat non-uniform.
However, they are easily explained as the accretion disc is still undergoing a global evolution over the duration of the evaluated time-window ($\Delta T = | 3000 - 4000 |$ \rgc).
At the beginning ($T=3000$ \rgc), we find a mean of $\bar{\Psi}_\mathrm{B} \approx 6.5$, while at the end ($T=4000$ \rgc) we find a mean of $\bar{\Psi}_\mathrm{B} \approx 11.5$.

The last unexplained panels of Fig. \ref{fig:grmhd_M5_twoD_width} are two variations on the orbital velocity $\Omega = u^\phi / u^t$.
First, in panel ($l.e$ \& $r.e$), we investigate the ratio between the surface-average within the plasmoid contour ($\bar{\Omega}_\mathrm{in}$) with the surface-average for a shell directly outside the plasmoid contour ($\bar{\Omega}_\mathrm{shell}$).
The outer edge of the shell corresponds to one-and-a-half times the distance to the central O-point.
From this quantity we can gauge if the plasmoid moves with its surroundings ($\bar{\Omega}_\mathrm{in}/\bar{\Omega}_\mathrm{shell} = 1$) or disconnected from it ($\bar{\Omega}_\mathrm{in}/\bar{\Omega}_\mathrm{shell} \neq 1$).
From the distributions, we find that the mean is consistent with $\bar{\Omega}_\mathrm{in}/\bar{\Omega}_\mathrm{shell} = 1$, but their is also a significant variance indicating that the plasmoid can move twice as fast or slow with respect to its direct environment.
This can potentially be interpreted in a number of ways, which includes, e.g., that the plasmoid seems to be dynamically disconnected from the accretion disc in its direct surroundings (potentially driven by a plasma-instability).

Second, in panels ($l.j$ \& $r.j$), we evaluate the ratio of $\bar{\Omega}_\mathrm{in}$ divided by the Keplerian circular orbital velocity in the equatorial plane which is defined as $\Omega_\mathrm{K} = (x^{3/2} + a_\ast)^{-1}$ with $x$ the cylindrical radius (corresponding the horizontal axis in Figs. \ref{fig:grmhd_overview_iM5} and \ref{fig:grmhd_overview_rM5}) and $a_\ast=0.9375$ the black hole spin parameter.
It has been established that MAD discs are sub-Keplerian \citep{igumenshchev08,porth21} which explains the mean of $\bar{\Omega}_\mathrm{in}/\Omega_\mathrm{K}\approx0.8$.
Nevertheless, the broad distribution with $0.1 \lesssim \bar{\Omega}_\mathrm{in}/\Omega_\mathrm{K} \lesssim 1.3$ indicates the potential for plasmoids to be super- or sub-Keplerian, which has interesting observational implications.
However, one still has to take into account that our estimate of the Keplerian orbital velocity is somewhat crude as the plasmoids have non-zero $u^\theta$ or $u^r$ velocities that break both the circular and equatorial assumption for $\Omega_\mathrm{K}$. 

Even though the distributions of {\tt iM3}, {\tt rM3}, {\tt iM4}, and {\tt rM4} are not explicitly shown, we have confirmed that the general trends described for {\tt iM5} and {\tt rM5} are consistent with the lower resolution simulations. The quantitative differences in plasmoid identication rate ($N_\mathrm{P}$) will, however, be outlined explicitly for all cases in section \ref{res:grmhd_plasmoid_and_horizon_timeseries}.

\subsubsection{Plasmoid distribution functions} \label{res:grmhd_plasmoid_distribution_function}

\begin{figure*}
  \centering
  \includegraphics[width=0.49\textwidth]{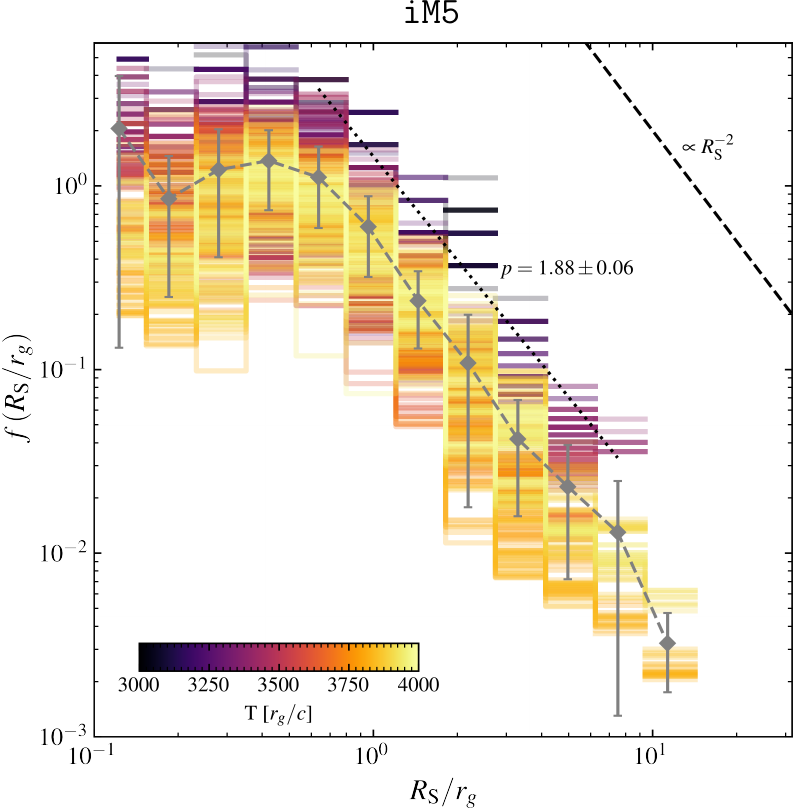}
  \hfill
  \includegraphics[width=0.49\textwidth]{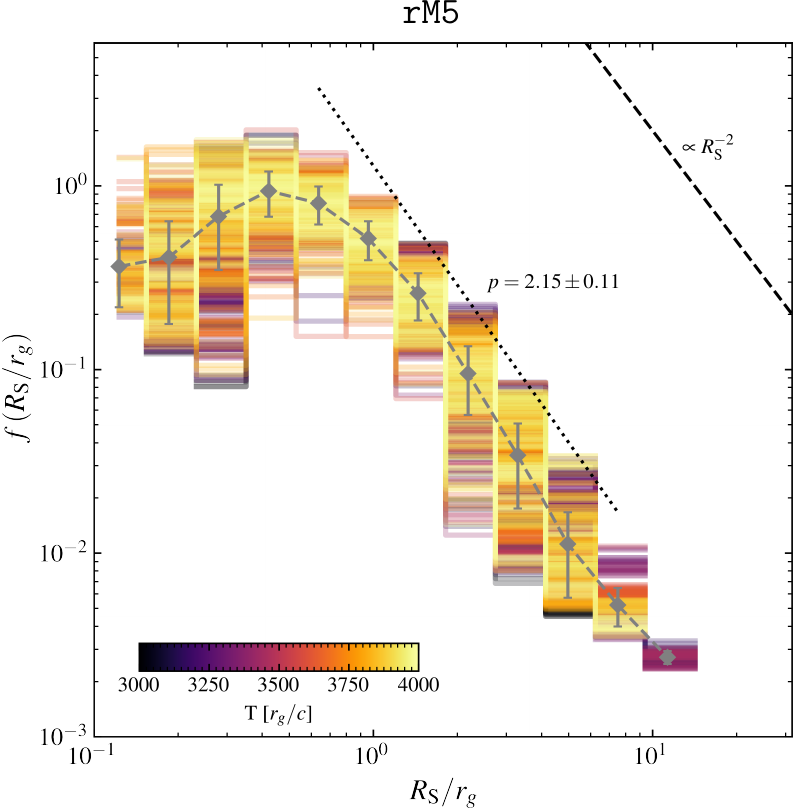}
  \caption{Probability density function $f (R_\mathrm{S} / r_g)$ of ``circular'' plasmoid radius $R_\mathrm{S}$ for both the high-resolution cases {\tt iM5} (\emph{left}) and {\tt rM5} (\emph{right}). All identification takes place within a circle of radius $R = 25 r_g$ and we evaluate a time-window of $T \in [3000,3001,\dots,3999,4000] \: r_g/c$. The rest of the description for Fig. \ref{fig:harris_pdf_width} is also applicable here, except now we utilize $R_\mathrm{S}$.
  }
  \label{fig:grmhd_i_and_rM5_pdf_RSwidth}
\end{figure*}

\begin{figure*}
  \centering
  \includegraphics[width=0.49\textwidth]{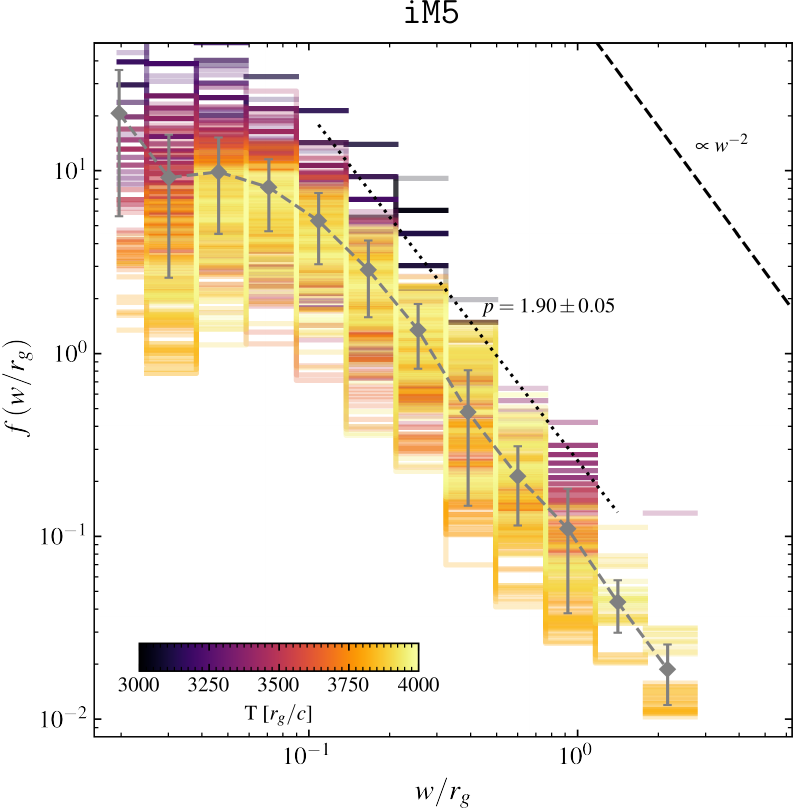}
  \hfill
  \includegraphics[width=0.49\textwidth]{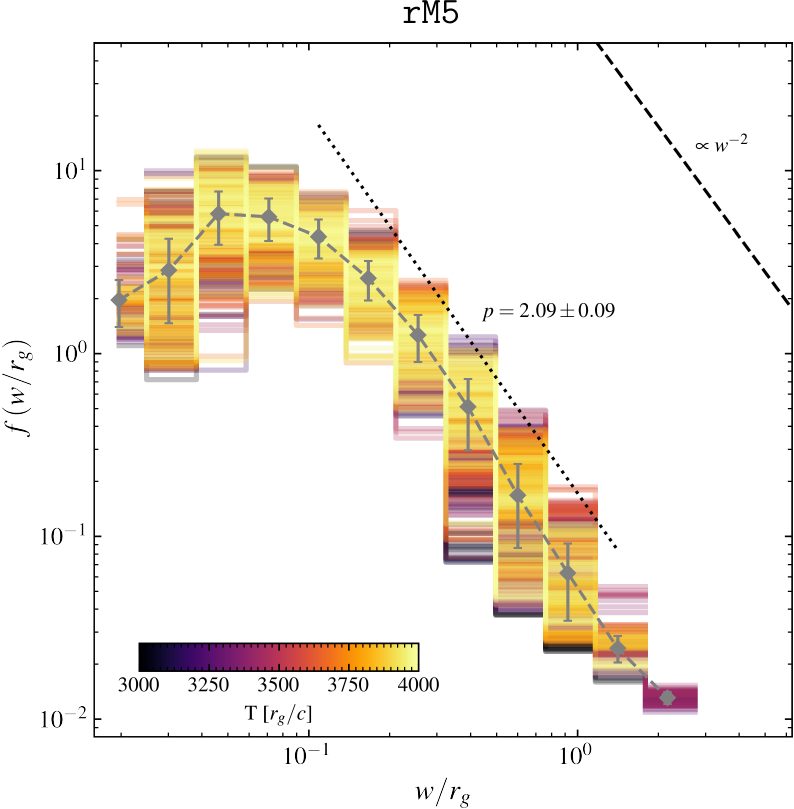}
  \caption{Probability density function $f (w / r_g)$ of plasmoid half-width $w$ for both the high-resolution cases {\tt iM5} (\emph{left}) and {\tt rM5} (\emph{right}). The rest of the description for Fig. \ref{fig:harris_pdf_width} is also applicable here. Note that the quantities here do not correctly take into account the spacetime curvature, which is the case for Fig. \ref{fig:grmhd_i_and_rM5_pdf_RSwidth}.
  }
  \label{fig:grmhd_i_and_rM5_pdf_width}
\end{figure*}


Figure \ref{fig:grmhd_i_and_rM5_pdf_RSwidth} displays the probability density function ($f$) of plasmoid radius $R_\mathrm{S}$, while Fig. \ref{fig:grmhd_i_and_rM5_pdf_width} displays the probability density function of plasmoid half-width $w$.
We show both distributions to illustrate the general-relativistic effects in Fig. \ref{fig:grmhd_i_and_rM5_pdf_RSwidth}, while Fig. \ref{fig:grmhd_i_and_rM5_pdf_width} is straight-forwardly compared with the Harris sheet's density function (in section \ref{res:harris_plasmoid_distribution_function}) and reflects the plasmoids shown in Figs. \ref{fig:grmhd_overview_iM5} and \ref{fig:grmhd_overview_rM5}. 
Interestingly, we recover the power law indices of $p=1.88\pm0.06$ ($p=1.90\pm0.05$) and $p=2.15\pm0.11$ ($p=2.09\pm0.09$) for {\tt iM5} and {\tt rM5} in Fig. \ref{fig:grmhd_i_and_rM5_pdf_RSwidth} (\ref{fig:grmhd_i_and_rM5_pdf_width}), respectively. 
These are similar to the results described in section \ref{res:harris_plasmoid_distribution_function}, which gives an indication that plasmoid formation is driven by the same principles even while taking into account the curvature of the spacetime.
So, even with the additional perturbations by the plasma instabilities outlined in section \ref{res:grmhd_general_evolution}, we still find scaling that is consistent with $p\approx2$.
While the onset of magnetic reconnection in the isolated Harris sheet simulations occurs somewhat spontaneously, in GRMHD it is subjected to global dynamics (such as the RTI and KHI) that trigger magnetic reconnection.
Although one clearly sees Harris-sheet-like structures forming in GRMHD, they also rapidly fall apart which interestingly does not affect the trends in the density functions.
One therefore concludes that the width distributions are robust features of reconnection, no matter how it is triggered.

In a way, the identification strategy we employ for the GRMHD simulations is more consistent with the aforementioned works that have outflowing boundary conditions as we stop identifying plasmoids when $r > 25$ \rg. 
Another ``outflowing'' boundary lies at the horizon but the vast majority of plasmoids moves outwards (in $\hat{r}$) in the jet-disc region.
Some plasmoids, typically associated with green contours, even move into the identification domain along the equatorial plane to then exit via the upper or lower identification boundaries.
Only a relatively minor fraction of plasmoids is accreted onto the BH and the majority of those are created in close proximity to the BH in the equatorial current sheet.

Close examination of Fig. \ref{fig:grmhd_i_and_rM5_pdf_RSwidth} yields that plasmoid radius goes all the way up to $R_\mathrm{S} \approx 10 r_\mathrm{g}$.
The Cartesian projection equivalent in Fig. \ref{fig:grmhd_i_and_rM5_pdf_width} yields a radius of $w \approx 2 \, r_\mathrm{g}$. 
These largest plasmoids are visible in Fig. \ref{fig:grmhd_overview_iM5}.
The smallest detected plasmoid radii correspond to $R_\mathrm{S} \approx 10^{-1} \, r_\mathrm{g}$ (and $w \approx 10^{-2} \, r_\mathrm{g}$).
Especially the largest plasmoids seem to be comparable in size to the `hot spots' that were used to interpret flares around Sgr A$^\ast$ \citep{gravity_jimenez20,wielgus22,vos22}.
From our simulations, we find that the plasmoids are of sufficient size to give a physical origin to these hot spots.
However, currently, we do not explicitly interpret their emission potential, but as plasmoids are typically hot ($p/\rho \gtrsim 1$) and magnetized ($\langle \bar{\sigma} \rangle \gtrsim 0.1$, as per Fig. \ref{fig:grmhd_M5_twoD_width}) they are likely to create a emission feature, albeit undetermined if predominantly thermal or non-thermal \citep{werner18,petropoulou18}.
Nevertheless, the occurrence rate of these large, and potentially bright, plasmoids is still quite low.
More specifically, for {\tt rM5}, plasmoids with radii $R_\mathrm{S} > 2.5 \, r_\mathrm{g}$ occur at least once and three times on average for all evaluated time instances (corresponding to $8.2\%$ of all identified plasmoids), while plasmoids with (Cartesian-projected) widths $w > 1 \, r_\mathrm{g}$ are much less common as they occur in only half ($51.4\%$) of the evaluated snapshots (corresponding to $1.8\%$ of all identified plasmoids).
This perceived discrepancy is partially due to the space-time curvature (not taken into account for $w$) and the mixing of plasmoid length and width for the $R_\mathrm{S}$ quantity.
For {\tt iM5}, the occurrence rates of at least one plasmoid passing the $R_\mathrm{S}$ and $w$ criteria are $57.7\%$ and $16.5\%$ (with $6.6\%$ and $2.3\%$ for all identified plasmoids over the entire time window), respectively.
Overall, if we take into account the much lower plasmoid counts for {\tt iM5}, we find that the percentage between the two cases are comparable, except for having at least one $w>1$ plasmoid per evaluated time.
This is well-explained, however, in section \ref{res:grmhd_plasmoid_and_horizon_timeseries}.

Lastly, we note that the power law gradient $p = - \mathrm{d} \log f / \mathrm{d} \log (R_\mathrm{S} / r_\mathrm{g})$ is less steep for {\tt iM5} than for {\tt rM5}.
We believe this is largely explained by the lower plasmoid number, but we also note that the colors indicate that at later times (more yellow) the plasmoid density function spans more radius (or width) bins and therefore lies slightly lower than at earlier times (dark purple to black).
This indicates there is some evolution in the density function as is confirmed in section \ref{res:grmhd_plasmoid_and_horizon_timeseries}.
For {\tt rM5}, we find a relatively consistent density function over time.
Next to a potential difference in evolution, we find that a singular linear relation (in log-log space) is not the best description of the downwards power law.
Even though close to $p\approx2$, there is a minor break visible and the gradient becomes shallower at $R_\mathrm{S}/r_\mathrm{g} \approx 4$.
As especially the larger plasmoid size bins contain more counts, this naturally pushes $p$ to slightly lower values for {\tt iM5}.
Nevertheless, it is interesting that {\tt rM5} indicates a somewhat steeper gradient with $p=2.15 \pm 0.11$.
However, combined with the points raised at the end of section \ref{res:harris_plasmoid_distribution_function}, we conclude that {\tt iM5} and {\tt rM5} are consistent with a power law with $p\approx2$ as more robust claims can not be made without further investigation.






\subsubsection{Timeseries of plasmoids and horizon penetrating fluxes} \label{res:grmhd_plasmoid_and_horizon_timeseries}

\begin{figure}
  \centering
  \includegraphics[width=\columnwidth]{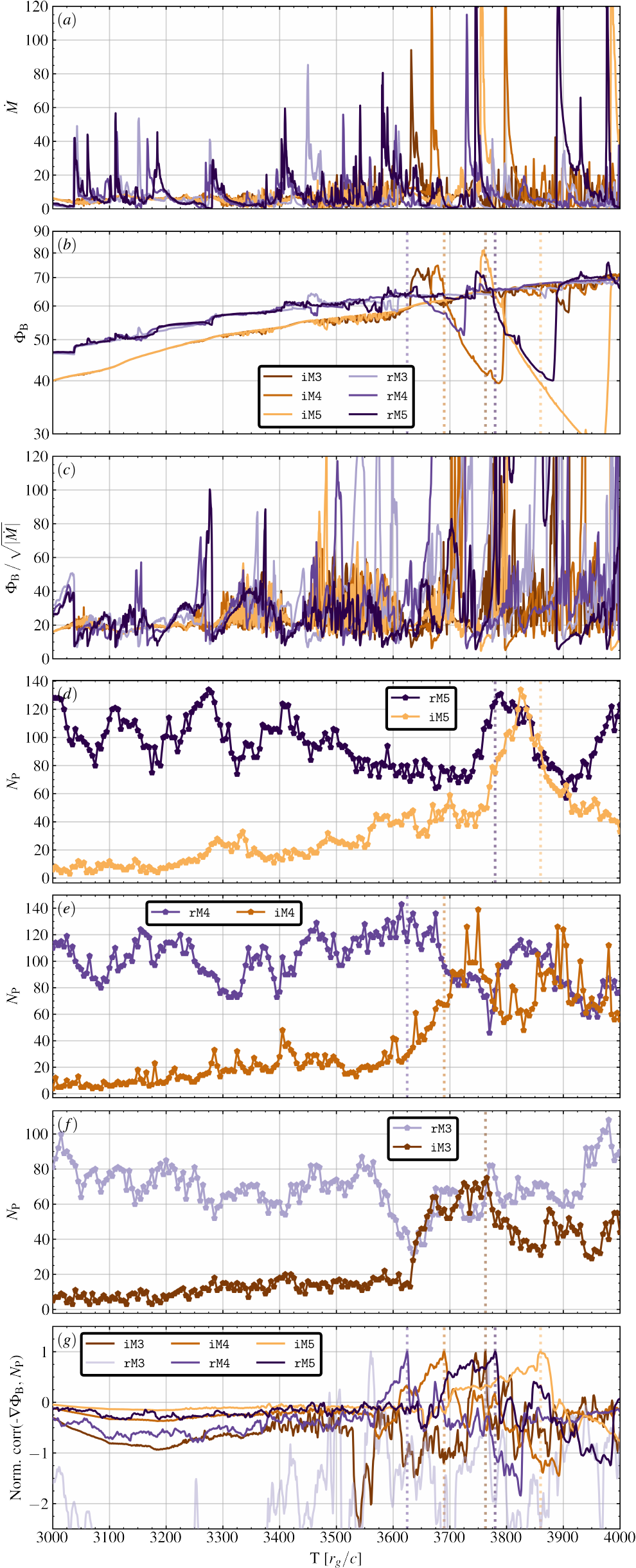}
  \vspace{-0.51cm}
  \caption{Timeseries of the mass accretion rate $\dot{M}$ (panel $a$), magnetic flux $\Phi_\textrm{B}$ ($b$), normalized magnetic flux $\phi = \Phi_\mathrm{B} / \sqrt{|\dot{M}|}$ ($c$), number of identified plasmoids $N_\textrm{Plasmoids}$ per simulation (panels $d$, $e$, and $f$), and normalized cross-correlation function (corr) between $-\nabla \Phi_\mathrm{B}$ and $N_\mathrm{P}$ ($g$). The fluxes are calculated at $2.5$ \rg. We display both the ideal ({\tt iM3}, {\tt iM4}, and {\tt iM5} in shades of \textit{orange}) and resistive ({\tt rM3}, {\tt rM4}, and {\tt rM5} in shades of \textit{purple}) GRMHD simulations. }
  \label{fig:grmhd_timeseries}
\end{figure}

Plasmoids form within the accretion disc and are mostly governed by their local plasma conditions.
Nevertheless, it would be interesting to see how these accretion disc probes connect dynamically to the central black hole.
Quantities that are typically calculated to assess this are the mass accretion rate ($\dot{M}$) and surface-penetrating magnetic flux ($\Phi_\mathrm{B}$) that are defined as \citep[in][]{porth19};
\begin{align}
  &\dot{M}         = -\int_{0}^{2 \pi} \int_{0}^{\pi} \rho u^r \sqrt{-g} \ d\theta \ d\phi , \, \label{eq:mdot} \\
  &\Phi_\mathrm{B} = \frac{1}{2}\int_{0}^{2 \pi} \int_{0}^{\pi} |^* F^{rt}| \sqrt{-g} \ d\theta \ d\phi . \, \label{eq:phib} 
\end{align}

MAD models are known to saturate in horizon-penetrating magnetic flux.
This implies that magnetic energy will be building up and will eventually be released in a sudden flux eruption that partly and temporarily halts the accretion flow onto the BH.
In two-dimensional simulations, the accretion flow will be stopped completely due to the constraining nature of the setup.
The parameter that is used to quantify this behavior is the so-called MAD parameter $\phi_\mathrm{BH} = \Phi_\mathrm{B} / \sqrt{\dot{M}}$, which corresponds to the normalized magnetic flux.
The MAD parameter saturates (in 3D) at $\phi_\mathrm{BH} \approx 15$\footnote{Note that this differs by a factor $\sqrt{4\pi}$ from works like \citet{tchekhovskoy11,tchekhovskoy12,mckinney12}. There, $\phi_\mathrm{BH}$ saturates at $\sim 50$.} \citep[cf.][]{yuan14}.
In our simulations, as shown in Fig. \ref{fig:grmhd_timeseries}, we find that $\phi$ occasionally rises to $\phi_\mathrm{BH} \sim 120$.
This is due to the confining nature of the 2D simulation, which allows for a greater accumulation of magnetic flux before an eruption.
It is consistent with behaviour found for simulations in \citet{ripperda20}.
As we used a different adiabatic index $\hat{\gamma}=13/9$ (vs. $\hat{\gamma}=4/3$ for \citealt{ripperda20}), we have a thicker disc at initialization which allows for greater accumulation of magnetic flux.

The middle to lower panels ($d$-$f$) of Fig. \ref{fig:grmhd_timeseries} display the number of identified plasmoids $N_\mathrm{P}$ per simulation.
While not shown explicitly in the figure, we confirm that plasmoids for either polarity (i.e., purple and green contours in Figs. \ref{fig:grmhd_overview_iM5} and \ref{fig:grmhd_overview_rM5}) are equally abundant.
As we already indicated (in section \ref{res:grmhd_plasmoid_statistics}), a significantly lower number of plasmoids is detected for the ideal simulations than for the resistive ones where a factor $2-10$ difference (in $N_\mathrm{P}$) is common.
The mechanism that triggers plasmoid formation, via the tearing instability, is not well-defined in ideal simulation and, more specifically, resolution dependent ($\propto \Delta x ^ 2$, with $\Delta x$ being the cell-size).
This implies that numerical resistivity ($\eta_\mathrm{ide} = \eta_\mathrm{num}$) is lower close to the black hole then further away due to the MKS coordinate system and is significantly smaller than $\eta_\mathrm{ris}=\eta$ ($\eta_\mathrm{num} << \eta$).
Overall, the tearing-instability is triggered less often, due to the relatively lower resistivity, and less reliably as it is determined by (stochastic) numerical effects.
Visually, the ideal simulations are significantly calmer, which is explained by the suppression of the MRI after the initial few thousand time-steps. 
Starting from $T\approx3700$ \rgc, however, a sudden increase in plasmoid formation rate is visible, which roughly corresponds to the state shown in Fig. \ref{fig:grmhd_overview_iM5} for {\tt iM5}.
After this ``flaring'' event, the rate at which plasmoids are created is somewhat increased (except for {\tt iM5}).

The resistive simulations possess a surprisingly constant number of plasmoids ($N_\mathrm{P}$), indicating a steady rate of plasmoid formation.
As the MRI is also suppressed for the resistive simulations, we can assume that the tearing instability is a sufficient perturbation in itself to keep plasmoid formation up.
To get an indication of how the flux eruptions could contribute to this process, we verified if there are significant changes in the $M_{\Delta T} \equiv \sigma_{\Delta T} / \mu_{\Delta T}$ \citep[see][and description therein]{eht22sgrav}, with $\sigma_{\Delta T}$ and $\mu_{\Delta T}$ being the standard deviation and mean, respectively, over a time-interval $\Delta T = |3000 - 4000|$ \rgc.
We calculated the modulation index for both the accretion rate $\dot{M}$ and magnetic flux $\Phi_\mathrm{B}$ that penetrate the spherical shell at $r=2.5$ \rg. 
The modulation indices for our simulation are listed in Table \ref{tab:GRMHDrms}.
There seems to be little difference between $M_{\dot{M}}$ or $M_{\Psi_\mathrm{B}}$ for the ideal and resistive simulations.
This is surprising as $N_\mathrm{P}$ indicates a more turbulent disc for the resistive cases, as this would give rise to the greater plasmoid count.
Nevertheless, one is not able to ascertain this directly from the shell-penetrating fluxes.
Another consequence of setting a fixed resistivity is that there is a fixed length-scale (i.e., width of the current sheet) that determines when the tearing instability is triggered.
When this length-scale is sufficiently resolved, one finds consistent results starting from a certain critical resolution and upwards.
It is therefore interesting that we see this being verified in the panels (d)-(f).
For {\tt rM3}, the lowest resolution case, we find that the mean plasmoid count $\langle N_\mathrm{P} \rangle \sim 75$.
While for the higher resolution cases {\tt rM4} and {\tt rM5}, we find $\langle N_\mathrm{P} \rangle \sim 100$.
As we find converging plasmoid numbers for both resolution cases, we conclude that the current sheet width (set by $\eta=5\times10^{-5}$), within the $25$ \rg domain, is fully resolved starting from a resolution of $4096^2$.

In the last panel ($g$), we cross-correlate the plasmoid number ($N_\mathrm{P}$) with the negative gradient of the magnetic shell-penetrating flux ($-\nabla \Psi_\mathrm{B}$) and find a positive relation for most cases.
Except for {\tt rM3}, which is the uncorrelated component on the background (in lightest purple), we find a clear correlation between the most prominent peak in $N_\mathrm{P}$ and a decrease in $\Psi_\mathrm{B}$.
The maxima of the correlation function coincides with beginning of a drop in the magnetic flux function and are denoted by vertical dashed line in their corresponding panels.
This is a consistent trend as long as one has a clear flux eruption, which also explains the uncorrelated {\tt rM3} results as there is no clear decrease in $\Phi_\mathrm{B}$ present.
For {\tt iM5} at $T \approx 3780$ \rgc, the flux eruption is rather large as is indicated by the decrease in $\Phi_\mathrm{B}$, which has pushed the maximum corr($-\nabla \Phi_\mathrm{B}, N_\mathrm{P}$) further to the right.
Just before the flux eruption, we find that an increase (of several factors) in $\Phi_\mathrm{B}$ after which it will start to drop.
The positive correlation is naturally explained by the fact that the flux eruption, that is accompanied by the temporary halting of the accretion flow, is a significant perturbation to the accretion disc that is able to initiate reconnection in numerous places.
Even though this general picture applies, we find that the dynamics are likely also stochastic in nature as the {\tt rM4} case displays different behavior with a drop in $N_\mathrm{P}$ directly after the flux eruption.
This is in part explained by our identification strategy which only identifies plasmoids within $25$ \rg and as the disc has receded during the flux eruption the domain in which plasmoids can form also shrinks and effectively delays the peak in $N_\mathrm{P}$.
Additionally, the shell-penetrating magnetic flux ($\Phi_\mathrm{B}$) only shows a relatively minor depression which indicates a relatively minor flux eruption and subsequent perturbation of the disc structure.
So, in short, one can expect a reaction in the plasmoid formation rate following a flux eruption, which tends to increase the plasmoid count as it perturb the disc triggering reconnection. 

\begin{table}
    \centering
    \begin{tabular}{ ccccccc }  \toprule
      \emph{Name} & $\mu_{\dot{M}}$ & $\sigma_{\dot{M}}$ & $M_{\dot{M}}$ & $\mu_{\Phi_\mathrm{B}}$ & $\sigma_{\Phi_\mathrm{B}}$ & $M_{\Phi_\mathrm{B}}$ \\ \midrule
        \tt{iM3} & 6.86  & 7.05  & 1.03 & 55.99 & 8.87  & 0.16 \\  
        \tt{iM4} & 7.36  & 11.46 & 1.56 & 54.19 & 8.93  & 0.16 \\  
        \tt{iM5} & 7.29  & 17.35 & 2.38 & 50.45 & 10.17 & 0.2  \\  
        \tt{rM3} & 6.34  & 8.16  & 1.29 & 59.38 & 6.27  & 0.11 \\  
        \tt{rM4} & 7.29  & 9.23  & 1.27 & 59.37 & 6.55  & 0.11 \\  
        \tt{rM5} & 10.78 & 17.28 & 1.6  & 57.79 & 7.53  & 0.13 \\ \bottomrule
    \end{tabular}
    \caption{The modulation index $M_\mathrm{Q} \equiv \sigma_\mathrm{Q} / \mu_\mathrm{Q}$ with $\sigma_\mathrm{Q}$ and $\mu_\mathrm{Q}$ denoting the standard deviation and mean of quantity $\mathrm{Q} \in \{\dot{M}, \Phi_\mathrm{B}\}$. This index gives a measure of the variability in the simulations' timeseries.}
    \label{tab:GRMHDrms}
\end{table}

\section{Discussion} \label{disc:discussion}
In this section, we will discuss our results following in the context of earlier works following four main points; (i) direct comparison to GRMHD-related plasmoid detection methods, (ii) specifics from our simulation library, (iii) implication for three-dimensional (3D) results, (iv) effects of resistivity, and (v) a discussion of the flaring potential of plasmoids.

\textbf{GRMHD plasmoid detection.}
Comparison with earlier works that have identified plasmoid structures in GRMHD \citep{nathanail20,jiang23} suggests that the approach outlined in this work finds $5-10\times$ more plasmoids.
Both aforementioned works utilize the Bernoulli factor ($Be = -hu_t$) as underlying identification medium and use a canny-edge detection algorithm on a Gaussian blurred segment (as provided by the {\tt scikit} Python package).
We have made initial attempt with this proposed method but we did not reach the desired efficacy or fidelity, which started the development of algorithm outlined in this work.
Overall, we typically find $5-10\times$ more plasmoids than the previously mentioned works, which are not all attributed to the detection method difference.
Other potential causes for the discrepancy can be the identification medium, resolution, simulation configuration, and the inherent differences between resistive and ideal GRMHD.
A number of these points will be discussed in detail in the following paragraphs.

Starting with the identification medium, which we took to be the magnetic flux function $\Psi_\mathrm{B}$ as it naturally identifies places with circular magnetic field structure.
When we compare this with using the Bernoulli factor $Be$, then it is clear from the results in this work that not all plasmoids are unbound as demonstrated in Fig. \ref{fig:grmhd_M5_twoD_width}.
One is likely to miss the plasmoids created in the equatorial plane as those tend to be bound \citep[as was also pointed out in][]{jiang23}.
There are also clear advantageous to using $Be$, because one can apply well-established image-recognition algorithms if one is able to increase the contrast (i.e., only show a limited color-range) to which the $Be$ lends itself well.
Nevertheless, this comes with the cost that one can only identify a subset of the plasmoid population.

\textbf{Simulation library.} When visually comparing our simulation to those of \citet{ripperda20}, with a highest resolution of $6144 \times 3072$ with respect to our $8192 \times 8192$, then we infer that the number of plasmoids does not differ significantly based on the presented figures, except perhaps at the smallest scales.
More importantly, one may even draw the conclusion that SANE simulations produce clearer and more abundant plasmoid structures.
\citet{nathanail20} utilizes an initial single dipolar loop up to intricate multi-polar initial magnetic field configurations with an evolution that can be described as SANE-like (with low $\phi_\mathrm{BH} \sim 2$).
Especially, the multi-polar configurations are expected to produce a lot of plasmoids, as is confirmed in their Fig. 6.
However, they do not show any statistics.
This is done, however, in \citet{jiang23} using the same methodology, but their configuration has a multi-polar initial magnetic field and evolves to be heavily magnetized (i.e., MAD-like).
The evolution is very chaotic and consistent with MAD but only relatively few plasmoids are visible indicating that the lower resolution (up to $4096 \times 2048$) and identification technique are likely to play a role.
It is important to note that those simulation were using ideal MHD, so we only compare it to the {\tt iM3}, {\tt iM4}, and {\tt iM5} cases. 
The differences between resistive and ideal GRMHD will be discussed in detail in section \ref{conc:conclusion}.

\textbf{3D.}
How applicable are 2D results to a 3D reality?
A number of arguments come into play here.
First, the plasmoids in our simulations describe predominantly elliptical (close to circular) structures and have long merging chains.
This is in part explained by the confined nature of the 2D simulations. 
As, due to this confining nature, plasmoids have a greater probability to interact and merge, they are likely to become larger.
If one were to add an additional dimension (in $\hat{\phi}$), one significantly complicates the situation. 
First, the plasmoid morphology would change and gain the resemblance of a flux rope.
Second, the chance for interaction would decrease significantly as it is simply less likely to come across another flux rope.
Third, the definition of flux ropes coalesce is difficult as they likely merge in a single place but not in it's entirety.
These points are clearly demonstrated for the 3D equivalent of the Harris sheet as presented in, e.g., \citet{sironi14,cerutti14a}.
There, one finds complex behavior of and interaction between flux ropes that is partially due to the presence of the kink instability \citep[e.g.,][]{bromberg19,davelaar20} which is absent in axisymmetric simulations.
For high-resolution 3D GRMHD simulations, some evidence for the presence of plasmoids, or flux ropes, was presented in \citet{ripperda22}.
Nevertheless, the typical appearance and how much it stands out with respect to its environment is relatively unknown in 3D.

\textbf{Resistivity.}
In essence, setting a resistivity ($\eta$) allows for consistently resolving the underlying current sheets in the simulation, which in ideal (GR)MHD is ill-defined as it is numerically determined and therefore has a stochastic (and coordinate-dependent) component.
As is clearly outlined in \ref{res:grmhd_plasmoid_and_horizon_timeseries}, there is a clear discrepancy between the resistive and ideal simulations.
While the former has a relatively consistent plasmoids number $N_\mathrm{P} \sim 100$, the latter has a non-flaring count comparable to $N_\mathrm{P} \sim 10$.
So, even though these discrepancies were expected, they were not verified in regard to plasmoid count till now.
In part it can be a selection effect as the ideal simulation(s) entered a `quiet' phase with few perturbations to the disc structure, but it is interesting this does not happen for the resistive case.
However, in the light of recent finding by the Even Horizon Telescope Collaboration \citep{eht22sgrav}, where was pointed out that the (ideal) GRMHD simulations produce too variable emission signatures, one can draw the tentative conclusion that this is further worsened by the use of resistive MHD. 
Additionally, the physical interpretation of resistivity is that it is a proxy for kinetic effects, which are simulated self-consistently with PIC methods, but to assess what is the `correct' resistivity for our physical scenario is a non-trivial question \citep{selvi22}.
A rigorous (GRMHD) study including several resisitivities is therefore needed to make more robust claims, but this is rather computationally expensive as one needs to assure that the current sheets are well-resolved.

\textbf{Misidentification.}
For the approach outlined in this work, we are indiscriminate as to what properties the plasmoid should contain, except that it should correspond to a circular magnetic field geometry.
Even though this allows us to get a rather complete distribution, it is slightly sensitive to misclassifications, which happens mainly for overly dense region.
This is explained by the sensitivity of both the local extrema finder and the watershed algorithm -- even though it is only a minor deviation from the background, it is treated as if it is a plasmoid.
Overall, this happens only rarely.
What occurs more often is that plasmoids that are in close vicinity to each other are grouped as they have very similar $\Psi_\mathrm{B}$ signatures.
Except that this diminished the detected plasmoid count somewhat, it does not influence the surface-averaged quantities (and distributions) as they still probe the plasmoid structure.
As with all identification problems, the difficulty lies in finding a strategy that is able to bridge the various length-scales while not picking up on erroneous features.
This is largely determined by the blurring layer, which dictates the minimal size-scale to which one is sensitive and gives a handle on how much fine-structure one want to include.
As the large plasmoid tend to have a lot of fine-structure, one should apply a more aggressively blurring strategy. 
Even though our algorithm is accurate, it is by no means computationally fast to run, even despite parallelization attempts which should be intensified in the future.
At present, we do not give an exact number of misclassifications, but one is able to find a few in most snapshots while the vast majority (of $\mathcal{O}(100)$) is classified correctly.
The number of plasmoids that were not classified is also of $\mathcal{O}(1)$ and are predominantly caused by numerical instabilities in the contour-finding step of the algorithm that typically occur for relatively unclear `plasmoid' structures.

\textbf{Flaring potential.}
While we started this paper by talking about plasmoids as a potential connection to flares, it is nevertheless difficult to make direct emission interpretations. 
The main reason for this is that the emission properties of plasmoids in the BH accretion environment are still very unknown, especially as one would expect a significant non-thermal contribution.
The utilization of a thermal synchrotron proxy \citep[as in, e.g.,][]{porth19} would therefore likely give an unrealistic picture.
\citet{ripperda20} gave estimates of the synchrotron emission and its potential to explain flares and our estimates are of the same order.
Nevertheless, it would be beneficial to conduct a full radiative transfer study to accurately access the flaring potential of plasmoids including a non-thermal electron population or reconnection-dedicated description \citep{rowan17}.
This is an interesting avenue to pursue in the future, as it is possible to pin-point the plasmoid's location with the algorithm.

\section{Conclusions} \label{conc:conclusion}
We have been able to identify plasmoids in highly turbulent accretion disc surrounding SMBHs with a higher fidelity than has been achieved before, which allows for creating complete time-series and distributions with sufficient counts to assess the statistics. 
Additionally, we have also verified our methodology with a set of previously well-investigated Harris current sheet simulations and found they are consistent with finding from previous studies \citep{uzdensky10,loureiro12,huang12,sironi16}.
Interestingly, the scaling laws (outlined in sections \ref{res:harris_plasmoid_distribution_function} and \ref{res:grmhd_plasmoid_distribution_function}) for both the Harris sheet and the GRMHD simulation are very similar, which indicates that plasmoid formation in the more complex accretion disc environment does not differ fundamentally from the Harris sheet picture.
Using this newly developed algorithm has enabled us to better study the plasmoid population within MAD accretion discs, and has clearly laid bare discrepancies in plasmoid occurrence rates between ideal and resistive MHD that warrant further investigation with a more systemic study that includes other accretion scenario (e.g., SANEs).

The typical plasmoid in a MAD GRMHD simulation is equally dense and somewhat under-magnetized with respect to their surrounding, moves with its surroundings, and is likely to be unbound according to the Bernoulli criterion.
Nevertheless, this behavior describes the averages of distributions and does not describe the deviations which occur frequently.
Especially for the orbital velocities and boundedness of the plasmoids, one finds large spreads in the distributions.
This indicates that plasmoids can both occur as super- or sub-Keplerian features, which is currently still an active point of investigation within the community.
Magnetic saturation at the BH event horizon produces flux tubes in a violent event that (partially) pushes back the accretion flow for MAD simulations.
Even though this is one of the leading theories to explain flares around SMBHs \citep{dexter20,porth21}, they are established to orbit with strongly sub-Keplerian velocities, which is at odds with some observations.
The formation of plasmoids is, therefore, still a strong candidate for explaining both Keplerian \citep{gravity_plasmoids_18,gravity_baubock20} and super-Keplerian \citep{matsumoto20} near-infra-red observation of flares around Sgr A$^\ast$.
More specifically, we regularly recover plasmoid sizes that are comparable to the hot spots that were used to interpret flares at both NIR- and mm-wavelengths \citep{gravity_baubock20,gravity_jimenez20,wielgus22b,vos22}. 
Also, as we outlined in section \ref{res:grmhd_plasmoid_and_horizon_timeseries}, flux eruptions (corresponding with a decrease in horizon-penetrating magnetic flux $\Phi_\mathrm{B}$) and plasmoid formation are likely strongly correlated with one another, indicating that flux eruptions act as an instigator of magnetic reconnection.
Both the flux tube and plasmoid (or flux rope) pictures do therefore not have to be mutually exclusive, but rather have a complementary co-existence. 

Lastly, we would like to point out that the identification algorithm is much more universally applicable as its function can be well-characterised as a `closed contour-detector around local extrema'.
So, in the future, we are planning to apply our methodology to mapping 3D structure of plasmoids and/or flux tubes for accretion onto SMBHs.
It would also lend itself well to other MHD or PIC identification applications, such as shearing or turbulent box simulations.

\section*{Acknowledgements}
We thank Bart Ripperda, Jordy Davelaar, Fiorenze Stoppa, Alejandra Jimenez-Rosales, and Aristomenis Yfantis for the helpful discussions and comments on the manuscript.
JV acknowledges support from the Dutch Research Council (NWO) supercomputing grant No. 2021.013.
MM acknowledges support by the NWO grant No. OCENW.KLEIN.113 and support by the NWO Science Athena Award.
BC acknowledges the European Research Council (ERC) under the European Union’s Horizon 2020 research and innovation program (Grant Agreement No. 863412).
HO acknowledges funding from Radboud University through a Virtual Institute of Accretion (VIA) postdoctoral fellowship from the Netherlands Research School for Astronomy (NOVA).

Software used to create the results in this work; \bhac \citep{porth17,olivares19}, {\tt Python} \citep{vanrossum09python}, {\tt NumPy} \citep{harris20numpy}, {\tt matplotlib} \citep{hunter07matplotlib}, \newline {\tt GNU parallel} \citep{tange22}

\section*{Data Availability}
The data used for this work will be shared following a reasonable request to the authors.
The plasmoid identification algorithm will be made publicly available in following repository; \href{https://github.com/JesseVos/Plasmoid\_Finder}{https://github.com/JesseVos/Plasmoid\_Finder}, in due time, but will also be shared following a reasonable request before that time.
 

\bibliographystyle{mnras}
\bibliography{references}


 
\appendix

\section{Resolution Convergence for GRMHD simulations} \label{app:resolution_convergence}
We have performed our simulations at three resolution levels, from lowest to highest; $2048 \times 2048$ ({\tt iM3} \& {\tt rM3}), $4096 \times 4096$ ({\tt iM4} \& {\tt rM4}), and $8192 \times 8192$ ({\tt iM5} \& {\tt rM5}).
These correspond to the third, fourth, and fifth AMR level, which we will use for referance. 
In principle, the current sheet are well-resolved starting from the fourth level, which is consistent with the plasmoid number findings in Fig. \ref{fig:grmhd_timeseries}.
Nevertheless, it is important to note that only relatively short periods, of $1000$ \rgc, have been run at the fourth and fifth level.
These simulation have been started from the third level snapshot at $2900$ \rgc, but then the resolution is increased up to the desired level.
After a period where the simulation adapts to the new resolution level, we start evaluating the window $T \in [3000,4000]$ \rgc.
Next to the analysis described in the main text, it would be interesting to see how the structure changes as a function of resolution level.
Therefore, we have calculated the time-averaged profiles over the aforementioned time-window for \psib, density $\rho$, and magnetization $\sigma$ for all cases where we are especially interested in the difference with respect to the highest resolution case.

\begin{figure}
    \centering
    \includegraphics[height=0.425\textheight]{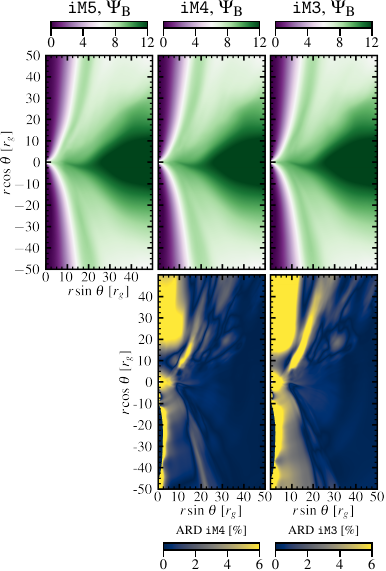}
    \caption{Time-averaged magnetic flux function \psib over time interval $T \in [3000,4000]$ \rgc for the {\tt iM5} (top left), {\tt iM4} (top middle), and {\tt iM3} (top right) cases. The bottom panels show the absolute relative difference (ARD) between the {\tt iM5} and the cases in the panels above. }
    \label{fig:appiMpsi}
\end{figure}

\begin{figure}
    \centering
    \includegraphics[height=0.425\textheight]{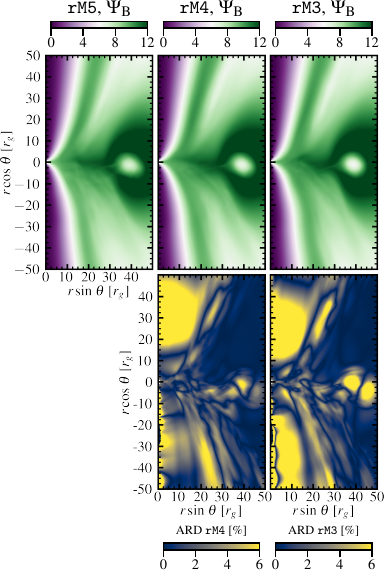}
    \caption{The description of Fig. \ref{fig:appiMpsi} applies here as well, except here we show the resistive cases; {\tt rM5}, {\tt rM4}, and {\tt rM3}.}
    \label{fig:apprMpsi}
\end{figure}

Figures \ref{fig:appiMpsi} and \ref{fig:apprMpsi} display the magnetic flux function \psib and the absolute relative difference (ARD) between the various resolution levels of the ideal and resistive simulations.
Starting with the ideal simulations, we predominantly find differences in the jet regions.
The inner jet region, near the axis, is dominated by numerical floor violations and does therefore not have a physical origin.
In the jet sheath, the transition regions between disc and jet, we do find most of the activity and difference between the various resolution levels.
Interestingly, for the ideal cases, most of the variability occurs in the upper ($x > 0$ \rg) region, while the bottom jet-sheath shows little variability.
This once again confirms that the evaluated ideal case is atypically quiet.
For the resistive cases, we find a similar structure with the highest values within the disc equatorial plane that then drops off the further you move away.
However, the jet sheath does have relatively higher flux function values.
The most striking difference is the gigantic plasmoid that lies at $x\approx45$ \rg on the equatorial plane. 
This is likely a remnant of the initial poloidal loop at initialization.  
Overall, we find much variability and differences between the various resolution cases, which indicates more activity overall (as was established throughout the main text). 
It is also important to note that the maximal differences are $\gtrsim 6 \%$, which is quite small and reasonable when compared to the other quantities.
This once again confirm that the flux function is a very suitable choice in identification medium as it is not very variable, which makes identification difficult.

\begin{figure}
    \centering
    \includegraphics[height=0.44\textheight]{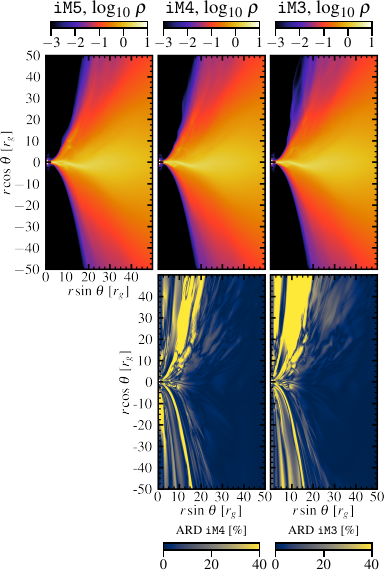}
    \caption{The description of Fig. \ref{fig:appiMpsi} applies here as well, except here we show the density $\rho$.}
    \label{fig:appiMrho}
\end{figure}

\begin{figure}
    \centering
    \includegraphics[height=0.44\textheight]{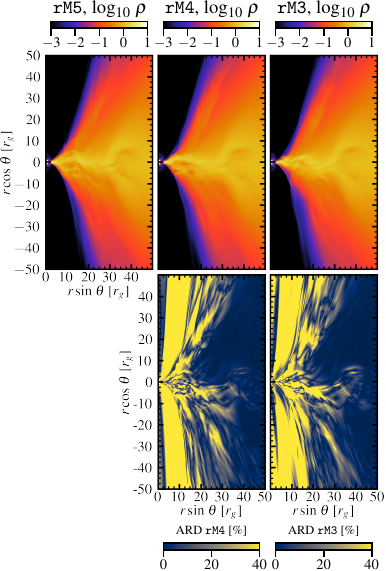}
    \caption{The description of Fig. \ref{fig:appiMpsi} applies here as well, except here we show the density $\rho$ for the resistive cases; {\tt rM5}, {\tt rM4}, and {\tt rM3}.}
    \label{fig:apprMrho}
\end{figure}

That leaves the time-averaged density ($\rho$) results in Figs. \ref{fig:appiMrho} and \ref{fig:apprMrho}.
Before we start the discussion, we would like to point out that the ARDs go up to $20\%$, which signifies that the differences are significantly larger.
This is in part due to the nature of the density itself as it tends to be small.
Nevertheless, we find that the results are consistent with what is shown for the \psib maps.
For the resistive cases, we find a lot of activity in both the equatorial plane (up to and concentrated around the giant plasmoid) and the jet sheath.
When compared to the ideal cases, we find that especially the equatorial current sheet activity is low.
This further outlines the clear differences between the ideal and resistive cases.
Nevertheless, we should be wary to take this as a general result, as it could very well be subject to selection effects.
We already noted in the past that the ideal case comes across as atypically quiet, which may have been the result of an unfortunate coincidence in time window.
Additionally, it is likely coincidental that a large equatorial plasmoid was created for the resistive cases, which has probably enhanced the resistive simulation's variability.
So, in the future, it would be interesting to undertake a more systematic study of resistive GRMHD simulations to see if the equatorial plasmoid is a common occurrence. 
Nevertheless, it is likely related to the confining nature of 2D simulations, as we commented before.

\bsp	
\label{lastpage}
\end{document}